\documentclass[aps,pra,showpacs,twocolumn,superscriptaddress]{revtex4-1}
\usepackage[colorlinks,linkcolor=blue,anchorcolor=blue,citecolor=blue,urlcolor=blue]{hyperref}
\usepackage{amsmath}
\usepackage[bb=boondox]{mathalfa}
\usepackage{bbm}
\usepackage{graphicx}
\usepackage{epstopdf}
\usepackage{physics}
\usepackage{algorithm}
\usepackage{algpseudocode}
\usepackage{amsfonts}
\usepackage{subfigure}
\usepackage{float}
\allowdisplaybreaks
\usepackage{color}
\newcommand*{\blue}{\textcolor{black}}

\begin{document}
	
	\title{Jenga-Krotov algorithm: Efficient compilation of multi-qubit gates\\for exchange-only qubits}
	
	\author{Jiahao Wu}
	\email{These authors contributed equally to this work.}
	\affiliation{Department of Physics, City University of Hong Kong, Tat Chee Avenue, Kowloon, Hong Kong SAR, China}
	\affiliation{City University of Hong Kong Shenzhen Research Institute, Shenzhen, Guangdong 518057, China}
	\affiliation{Quantum Science Center of Guangdong-Hong Kong-Macao Greater Bay Area, Shenzhen, Guangdong 518045, China}
	\author{Guanjie He}
	\email{These authors contributed equally to this work.}
	\affiliation{Department of Physics, City University of Hong Kong, Tat Chee Avenue, Kowloon, Hong Kong SAR, China} 
	\author{Wenyuan Zhuo}
	\affiliation{Department of Physics, City University of Hong Kong, Tat Chee Avenue, Kowloon, Hong Kong SAR, China} 	\affiliation{City University of Hong Kong Shenzhen Research Institute, Shenzhen, Guangdong 518057, China}
	\author{Quan Fu}
	\affiliation{School of Physics and Technology, Wuhan University, Wuhan 430072, China}
	\affiliation{Department of Physics, City University of Hong Kong, Tat Chee Avenue, Kowloon, Hong Kong SAR, China}
	\author{Xin Wang}
	\email{x.wang@cityu.edu.hk}
	\affiliation{Department of Physics, City University of Hong Kong, Tat Chee Avenue, Kowloon, Hong Kong SAR, China} 	\affiliation{City University of Hong Kong Shenzhen Research Institute, Shenzhen, Guangdong 518057, China}

	\begin{abstract}
Exchange-only (EO) qubits, implemented in triple-quantum-dot systems, offer a compelling platform for scalable semiconductor-based quantum computing by enabling universal control through purely exchange interactions. While high-fidelity single- and two-qubit gates have been demonstrated, the synthesis of efficient multi-qubit operations—such as the Toffoli gate—remains a key bottleneck. Conventional gate decompositions into elementary operations lead to prohibitively long and error-prone pulse sequences, limiting practical deployment. In this work, we introduce a gradient-based optimization algorithm, Jenga-Krotov (JK), tailored to discover compact, high-fidelity EO gate sequences. Applying JK to the Toffoli gate, we reduce the number of required exchange unitaries from 216 (in \blue{direct} decomposition) to 92, and compress the time steps required from 162 to 50, all while maintaining target fidelity. Under realistic noise, the accumulated gate error from our optimized sequence is an order of magnitude lower than that of conventional approaches. \blue{We have also applied the JK algorithm to other multi-qubit gates and algorithm. For the Fredkin gate, it reduces the number of time steps from 200 to 104 and the number of exchange unitaries from 276 to 172. For the quantum Fourier transform, it compresses the sequence from 180 to 80 time steps and from 237 to 202 exchange unitaries.}
These results demonstrate that the JK algorithm is a general and scalable strategy for multi-qubit gate synthesis in EO architectures, potentially facilitating realization of multi-qubit algorithms on semiconductor platforms.
\end{abstract}

	\maketitle
	
	\section{Introduction}
Quantum computing promises exponential speedup for problems intractable on classical machines, ranging from cryptographic factoring to quantum simulation. Over the past two decades, significant progress in quantum system design, control, and coherent manipulation has brought this vision closer to reality, leading to the Noisy Intermediate-Scale Quantum (NISQ) era \cite{Preskill2018,DeLeon.21,Chen2023,AbuGhanem2024}. However, NISQ devices suffer from imperfect gate fidelities and limited error correction, posing major obstacles to scalable computation \cite{Leymann2020}. 
Among the various physical platforms under development, semiconductor-based spin qubits stand out for their scalability, long coherence times, and compatibility with industrial fabrication processes
\cite{Burkard2023,Loss1998,Friesen.03,Petta2005,Hanson.07,Maune.12,Zajac.16,Nichol2017,Chatterjee.21,Weinstein2023}.

Within this class, Exchange-Only (EO) qubits --- encoded in triple quantum dots \cite{Schroeer.07,Eng.15,Blumoff.22} --- enable universal quantum computation using only Heisenberg exchange interactions, without requiring electron spin resonance or magnetic field gradients \cite{DiVincenzo2000,Levy.02,Laird.10,Medford.13b,Ha.21}. By operating within Decoherence-Free Subspaces (DFS) \cite{Bacon2000}, EO qubits offer fast and efficient gate operations that can potentially approach sub-nanosecond timescales \cite{Petta2005,Hu2025}. These features allow more operations to be executed within the limited coherence window, making EO qubits an attractive candidate for implementing quantum algorithms in NISQ devices.

While many variants of EO qubits have been proposed, such as hybrid, resonant-exchange, and quadrupolar designs \cite{Shi.12,Cao.16,Russ.17,Thorgrimsson.17,Frees.19,Medford.13,Taylor.13,Doherty.13,Russ.18,Qiao2020,Sala.20}; this work focuses on the canonical implementation: a linear triple-dot array, each dot hosting a single electron. This configuration offers a clean and well-controlled platform. In particular, it provides an ideal testbed for high-fidelity multi-qubit gate synthesis --- a key step toward scalable EO-based quantum processors.

Efficient implementation of multi-qubit gates is a central challenge in scaling up quantum computing, particularly in platforms like EO qubits, where gate operations are restricted to nearest-neighbor exchange interactions. While single- and two-qubit gates have been extensively studied and optimized---for example, the CNOT sequence proposed by Fong and Wandzura remains the most efficient known construction for EO systems \cite{Bryan2011,Zeuch2016,Zeuch2020}---the synthesis of multi-qubit gates remains under-explored due to its inherent complexity.

To address this gap, we focus on the construction of a three-qubit gate --- the Toffoli gate --- as a prototypical benchmark for scalable multi-qubit gate design. The Toffoli gate is a widely used universal gate for classical reversible computation and plays a central role in quantum algorithms and fault-tolerant architectures \cite{AbuGhanem2025,Nielsen2010,Maslov2004}.
While it is theoretically possible to decompose any multi-qubit gate into sequences of single- and two-qubit gates, this strategy becomes inefficient in the EO qubit setting. For instance, the direct decomposition (later called ``DIR'') of a Toffoli gate into single-qubit gates and CNOTs (using the Fong-Wandzura sequence) results in 216 unitaries and 162 time steps \cite{Cruz2024} (cf. Sec.~\ref{subsec:directdecomp}), which significantly amplifies decoherence and resource overhead. Furthermore, numerical efforts using existing optimal control methods like GRAPE or standard Krotov encounter challenges: they either demand impractical computational time and resources, or frequently become trapped in local minima or stall on barren plateaus in the fidelity landscape. 
These considerations motivate the development of new optimization strategies for direct synthesis of multi-qubit gates, starting with the Toffoli gate as a foundational step toward more complex gate constructions in EO qubit architectures.

In this work, we address these challenges by introducing a new gradient-based numerical method --- the Jenga-Krotov (JK) algorithm --- which enables the efficient discovery of high-fidelity, low-depth pulse sequences for multi-qubit gates. While our benchmark application focuses on the Toffoli gate in a linearly coupled triple-EO-qubit system, the JK algorithm is general and broadly applicable to the compilation of arbitrary multi-qubit gates and alternative geometries (e.g., triangular configurations \cite{Setiawan2014,Acuna.24} within the EO qubit framework), while overcoming key limitations of previous optimal control methods.
	
	The remainder of this paper is organized as follows: In Section~\ref{sec:model}, we introduce the EO qubit
	and its encoding within a DFS, and extend this framework to a system of three EO qubits. Section~\ref{sec:methods} presents methods to decompose a multi-qubit gate, including DIR, Gradient Ascent Pulse Engineering (GRAPE), and the Krotov’s method. We further present details of our JK algorithm for efficient pulse sequence optimization. In Section~\ref{sec:result}, we apply the JK algorithm to construct a high-fidelity Toffoli gate and evaluate its performance under realistic noise. Finally, Section~\ref{sec:concl} summarizes our main findings and discusses the broader implications of the JK algorithm for scalable EO qubit systems.

	\section{Model}\label{sec:model}
	
	\subsection{EO qubit}
	The EO qubit considered here consists of a linear array of three spin-1/2 particles (labeled 1, 2, 3 ---  there are complications arising from the order of the labelling which we will discuss later), with only nearest-neighbor exchange interactions. The system is governed by the Hamiltonian:
	\begin{equation}
  \begin{aligned}
	\mathcal{H}&=\sum_{\langle i,j\rangle}\mathcal{H}_{i,j},\\
		\mathcal{H}_{i,j}&=J_{i,j}\left(t\right)\boldsymbol{S}_i\cdot\boldsymbol{S}_j,\qquad\left(i,j=1,2,3\right),
 \end{aligned}
 \end{equation}
	where $\boldsymbol{S}_i=\frac{1}{2}\boldsymbol{\sigma}_i$ is the spin of particle $i$, $\boldsymbol{\sigma}_i$ is the corresponding Pauli operator, and $J_{i,j}\left(t\right)$ is the time-dependent exchange coupling between particles $i$ and $j$, used to implement exchange unitaries, and $\langle i,j\rangle$ indicates nearest neighbors. 
	
	The system has eight eigenstates, denoted  $|1\rangle\ldots|8\rangle$. These can be expressed both in the spin basis and in terms of the complete set of commuting observables 
 $\{S_{\mathrm{tot}}^{(3)},S_{z,\mathrm{tot}}^{(3)},S_{1,2}\}$  (here, the labels 1, 2, 3 are in \emph{natual order}, which will be explained later) as follows \cite{Laird.10,Ladd.12,Gaudreau.12}:

	\begin{equation}
  \begin{aligned}
   &\left|1\right\rangle=\left|\frac{1}{2},\frac{1}{2},0\right\rangle=\frac{1}{\sqrt{2}}\left(\left|\uparrow\downarrow\uparrow\right\rangle-\left|\downarrow\uparrow\uparrow\right\rangle\right)\\
   &\left|2\right\rangle=\left|\frac{1}{2},-\frac{1}{2},0\right\rangle=\frac{1}{\sqrt{2}}\left(\left|\uparrow\downarrow\downarrow\right\rangle-\left|\downarrow\uparrow\downarrow\right\rangle\right)\\
   &\left|3\right\rangle=\left|\frac{1}{2},\frac{1}{2},1\right\rangle=\sqrt{\frac{2}{3}}\left|\uparrow\uparrow\downarrow\right\rangle-\frac{1}{\sqrt{6}}\left|\uparrow\downarrow\uparrow\right\rangle-\frac{1}{\sqrt{6}}\left|\downarrow\uparrow\uparrow\right\rangle\\
   &\left|4\right\rangle=\left|\frac{1}{2},-\frac{1}{2},1\right\rangle=\frac{1}{\sqrt{6}}\left|\uparrow\downarrow\downarrow\right\rangle+\frac{1}{\sqrt{6}}\left|\downarrow\uparrow\downarrow\right\rangle-\sqrt{\frac{2}{3}}\left|\downarrow\downarrow\uparrow\right\rangle\\
   &\left|5\right\rangle=\left|\frac{3}{2},\frac{3}{2},1\right\rangle=\left|\uparrow\uparrow\uparrow\right\rangle\\
   &\left|6\right\rangle=\left|\frac{3}{2},\frac{1}{2},1\right\rangle=\frac{1}{\sqrt{3}}\left(\left|\uparrow\uparrow\downarrow\right\rangle+\left|\uparrow\downarrow\uparrow\right\rangle+\left|\downarrow\uparrow\uparrow\right\rangle\right)\\
   &\left|7\right\rangle=\left|\frac{3}{2},-\frac{1}{2},1\right\rangle=\frac{1}{\sqrt{3}}\left(\left|\uparrow\downarrow\downarrow\right\rangle+\left|\downarrow\uparrow\downarrow\right\rangle+\left|\downarrow\downarrow\uparrow\right\rangle\right)\\
   &\left|8\right\rangle=\left|\frac{3}{2},-\frac{3}{2},1\right\rangle=\left|\downarrow\downarrow\downarrow\right\rangle.
  \end{aligned}\label{eq:8eigenstates}
 \end{equation}
	
Here, the middle part of the equality chain displays the states in the form $\left|S_{\mathrm{tot}}^{(3)},S_{z,\mathrm{tot}}^{(3)},S_{1,2}\right\rangle$, where $S_{\mathrm{tot}}^{(3)}$ denotes the total spin of the three particles, $S_{z,\mathrm{tot}}^{(3)}$ is the total spin projection along the $z$-axis,  and $S_{1,2}$ represents the total spin of particles 1 and 2. the superscript ``$(3)$'' indicates the total of three spins. The right-hand side of Eq.~\eqref{eq:8eigenstates} shows the corresponding spin configurations; for example, $\left|\uparrow\downarrow\uparrow\right\rangle$ represents a state in which particle 1 is spin-up, particle 2 is spin-down, and particle 3 is spin-up.

The qubit state can be encoded in the $S_{z,\mathrm{tot}}^{(3)}=1/2$ subspace using states $\left|1\right\rangle$ and $\left|3\right\rangle$, which may leak into $\left|6\right\rangle$ under an inhomogeneous magnetic field; alternatively, one can use the $S_{z,\mathrm{tot}}^{(3)}=-1/2$ subspace with states $\left|2\right\rangle$ and $\left|4\right\rangle$, which may leak into $\left|7\right\rangle$ \cite{Ladd.12,Hickman2013,Andrews.19}. In this work, we assume a homogeneous external magnetic field and neglect the leaked states $\left|6\right\rangle$ and $\left|7\right\rangle$, therefore the qubit is defined entirely in the Decoherence-Free Subspace (DFS) \cite{Bryan2011}. In this paper, we take
$S_{z,\mathrm{tot}}^{(3)}=1/2$ therefore the qubit states are
$\left|\mathbb{0}\right\rangle=\left|1\right\rangle$ and $\left|\mathbb{1}\right\rangle=\left|3\right\rangle$, the label of the two qubit states (``$\mathbb{0}$'' and ``$\mathbb{1}$'') are identical to the corresponding $S_{1,2}$. We further note that once this convention is chosen, the three spins are no longer interchangeable. 

To facilitate later discussion, we follow a notation introduced in Ref.~\cite{Zeuch.14,Zeuch2016,Zeuch2020}.
In this notation, each spin is represented by the a bullet $\bullet$, and the groups of spins that related to chosen quantum numbers are enclosed in brackets (ovals in figures) labeled by their total spin. Therefore our $\left|\mathbb{0}\right\rangle=\left(\left(\bullet\bullet\right)_0\bullet\right)_{1/2}$ and $\left|\mathbb{1}\right\rangle=\left(\left(\bullet\bullet\right)_1\bullet\right)_{1/2}$, where the subscript for the inner brackets indicates $S_{1,2}$, while that for the outer brackets $S_{\mathrm{tot}}^{(3)}$.

\begin{figure}[htb]
	\centering
	\includegraphics[width=0.48\textwidth]{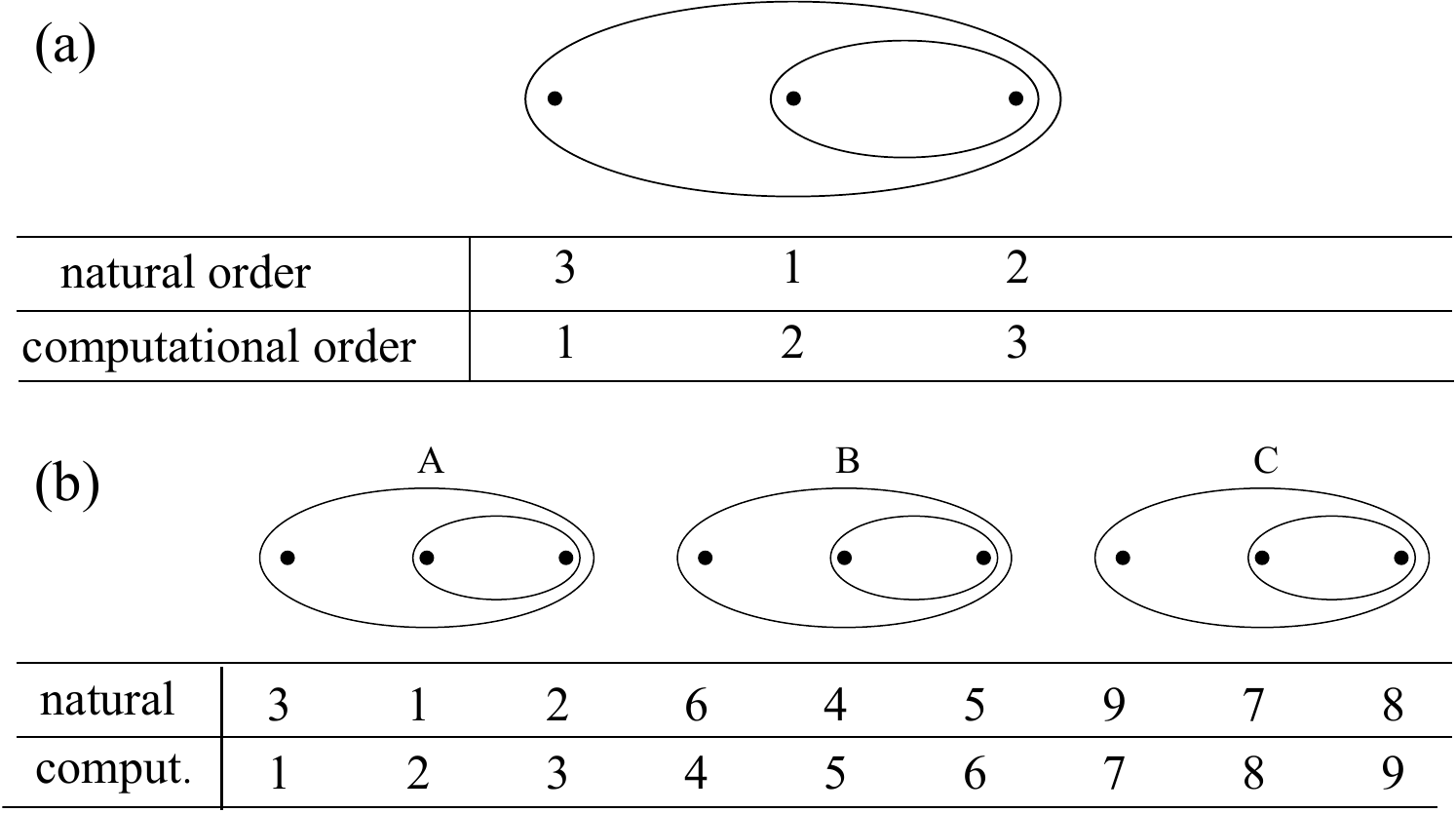}
	\caption{The bullet notation for (a) a single EO qubit, and (b) three EO qubits  labeled by A,B and C. The corresponding natural and computational orders are indicated.}
	\label{fig:bulletrepresentation}
\end{figure}

It is of utmost importance to discuss the arrangement of the three spins, which plays a critical role in our analysis. The conventional angular momentum coupling process inherently assumes a specific ordering: first, spin 1 and spin 2 are coupled, and then the resulting total spin is coupled to spin 3. We refer to this ordering as the \emph{natural order}. The eight states $|1\rangle \ldots |8\rangle$ introduced previously are constructed based on this natural order. On the other hand, when conducting quantum computation, we intuitively label spins from left to right (or top to bottom) with 1, 2, and 3. This is referred to as the \emph{computational order}. In some cases, such as 
$\left(\left(\bullet\bullet\right)_a\bullet\right)_{1/2}$, these two orders are the same, but in most situations that they are different. For example, in the original Fong-Wandzura sequence \cite{Bryan2011}, computational order (1,2,3,4,5,6) is actually (3,1,2,4,5,6) in natural order, namely $\left(\bullet\left(\bullet\bullet\right)_a\right)_{1/2}\left(\left(\bullet\bullet\right)_b\bullet\right)_{1/2}$.
This problem is even more pronounced when multiple EO qubits are coupled, as will be discussed later. In this work, we adopt the \emph{natural order} when angular momentum coupling is involved, and the \emph{computational order} when constructing Hamiltonians and evolution operators for quantum computation where the geometric ordering is more intuitive. The specific order being used will be clearly stated wherever ambiguity may arise. For clarity, Fig.~\ref{fig:bulletrepresentation} explicitly illustrates the correspondence between natural and computational orderings for both a single EO qubit and a three-qubit system.

To implement gates for an EO qubit, we adopt a piecewise-constant control protocol for 
$J_{i,j}(t)$, consistent with experimentally realizable square pulses and commonly used in optimal control simulations. The total evolution time $T$ is divided into segments. Each segment, with a fixed duration $\tau$ and constant exchange coupling $J_{i,j} = J$, is referred to as \emph{one} “exchange unitary,” a term we will use later for gate counting. To make the exchange unitaries explicit in our computational framework, we rewrite the Hamiltonian in \emph{computational order} as
\begin{equation}
\mathcal{H}_{i,j}^\mathrm{ex}=\frac{1}{4}\boldsymbol{\sigma}_i\cdot\boldsymbol{\sigma}_j-\frac{1}{4},\left(i,j=1,2,3\right).
\end{equation}
Defining $p = J\tau/\pi$, an exchange unitary between spins $i$ and $j$ is then given by
\begin{equation}
U_{i,j}^\mathrm{ex}\left(p\right)=e^{-i p\pi \mathcal{H}_{i,j}^\mathrm{ex}}.\label{eq:fundamentalblock}
\end{equation}

The task of implementing quantum algorithms and control protocols essentially reduces to finding a sequence of such exchange unitaries that realizes a target quantum gate. This naturally formulates a quantum optimal control problem, which can be addressed using either analytical or numerical methods.

\begin{figure*}[htb]
		\centering
		\includegraphics[width=1.6\columnwidth]{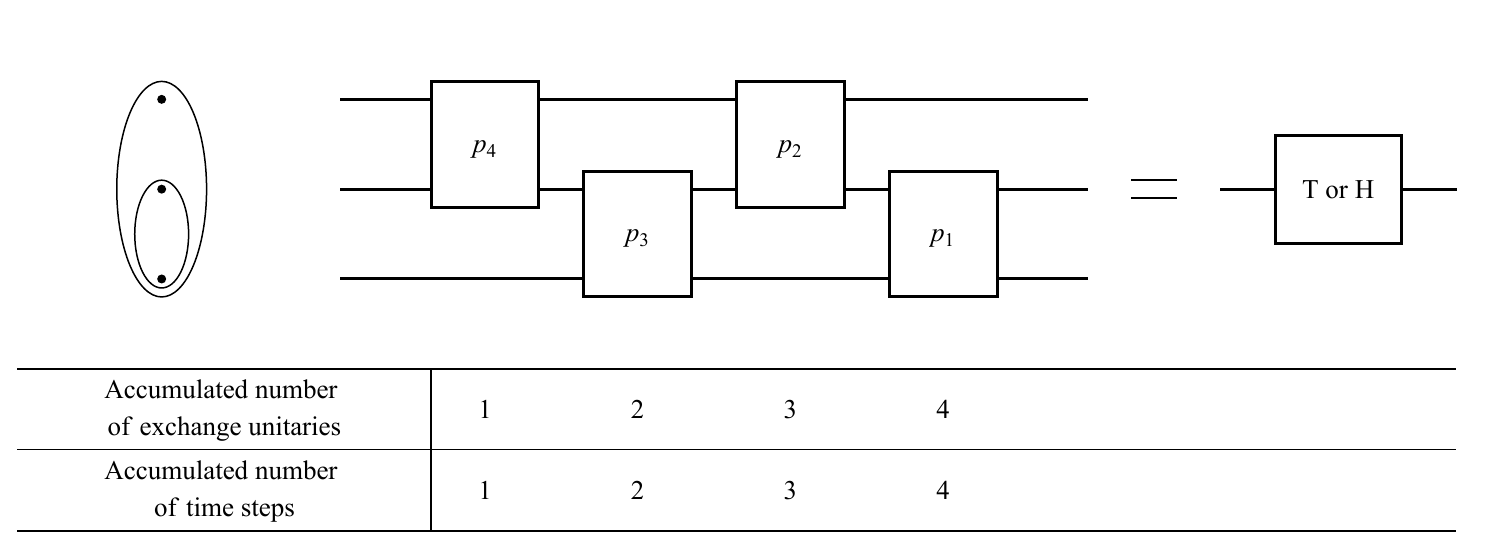}
		\caption{Quantum circuit showing direct decomposition of $T$ and $H$ gates, along with the accumulated number of exchange unitaries and number of time steps indicated. Each block corresponds to an exchange unitary as defined in Eq.~\eqref{eq:fundamentalblock}, with the $p$ value indicated inside the block.}
		\label{fig:THdecomposition}
	\end{figure*}

Here we give two explicit examples on single-qubit gates: a T gate  and a Hadamard gate (H). In this calculation, the computational order (1,2,3) is (3,1,2) in natural order, which means $\left|\mathbb{0}\right\rangle=\left(\bullet\left(\bullet\bullet\right)_0\right)_{1/2}$ and $\left|\mathbb{1}\right\rangle=\left(\bullet\left(\bullet\bullet\right)_1\right)_{1/2}$. Both of them can be decomposed into four exchange unitaries in a form of 
\begin{equation}
\mathrm{T}, \mathrm{H}=U_{23}\left(p_1\right)U_{12}\left(p_2\right)U_{23}\left(p_3\right)U_{12}\left(p_4\right).\label{eq:THdecomposition}
\end{equation}
The $p$ values are given in Table~\ref{tab:single-qubitgates}. The sequence in Eq.~\eqref{eq:THdecomposition} is also shown in Fig.~\ref{fig:THdecomposition} where the accumulated number of exchange unitaries and number of time steps are indicated. While this is merely a simple example, it provides a pedagogical starting point for later, much more complicated sequences.

	\begin{table}[htb]
		\centering
		\renewcommand{\arraystretch}{2} 
		\setlength{\tabcolsep}{10pt} 
		\begin{tabular}{|c|c c c c|}
			\hline
			Gate & $U_{23}\left(p_1\right)$ & $U_{12}\left(p_2\right)$ & $U_{23}\left(p_3\right)$ & $U_{12}\left(p_4\right)$ \\ \hline
			T & $\theta$ & $\frac{5}{4}-\theta$ & $\theta$ & $1-\theta$ \\ \hline
			H & $\phi_1-1$ & $1-\phi_2$ & $1+\phi_2$ & $2-\phi_1$ \\ \hline
		\end{tabular}
		\caption{compilation of a T gate  and a Hadamard (H) gate into four exchange unitaries.  $\phi_1=\arccos{\left(-1+\sqrt{2}+\frac{-2+\sqrt{2}}{\sqrt{3}}\right)}/{\pi},  \phi_2=\arccos{\left(\frac{-1+\sqrt{2}+\sqrt{6}}{3}\right)}/{\pi}$, and $\theta=1+\arccos{\left(\frac{2-\sqrt{2}+\sqrt{70+36\sqrt{2}}}{12}\right)}/\pi$.}
		\label{tab:single-qubitgates}
	\end{table}

\begin{figure*}[htb]
		\centering
		\includegraphics[width=2\columnwidth]{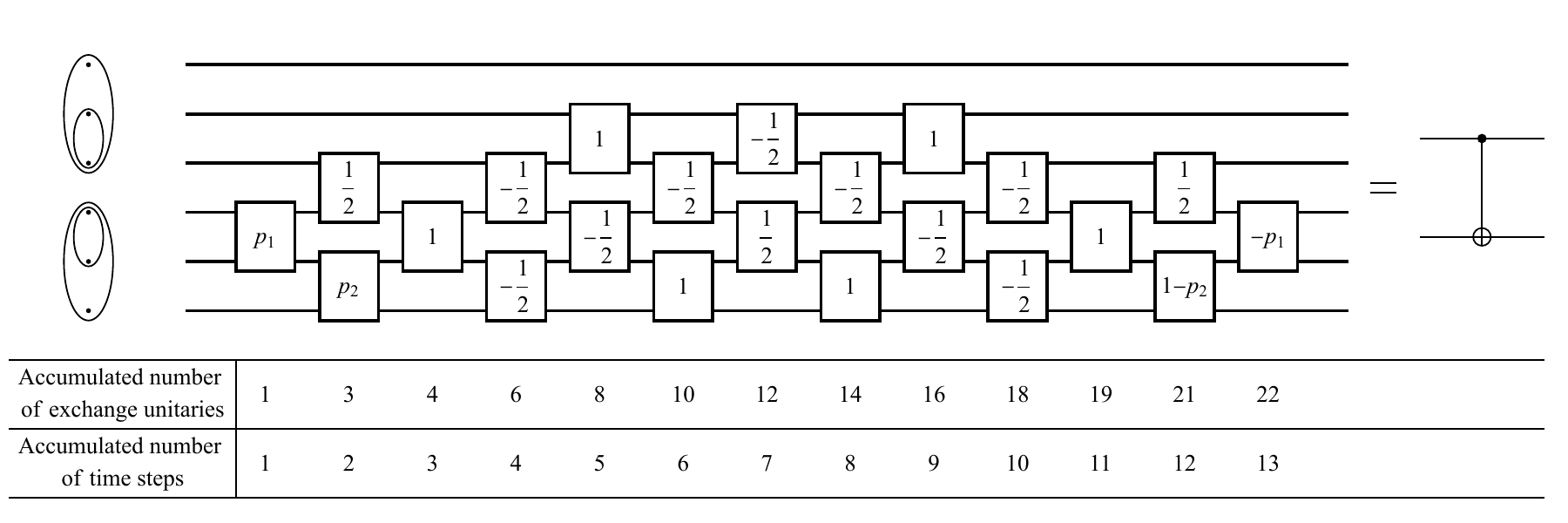}
		\caption{Quantum circuit showing the Fong-Wandzura sequence in the natural order (3,1,2,4,5,6) accomplishing a CNOT gate, along with the accumulated number of exchange unitaries and number of time steps indicated. Each block corresponds to an exchange unitary as defined in Eq.~\eqref{eq:fundamentalblock}, with the $p$ value indicated inside the block. Here $p_1=\arccos{\left(-1/\sqrt{3}\right)}/\pi$, $p_2=\arcsin{\left(1/3\right)}/\pi$.}
		\label{fig:312456FWsequence}
	\end{figure*}

\begin{figure*}[htb]
		\centering
		\includegraphics[width=2\columnwidth]{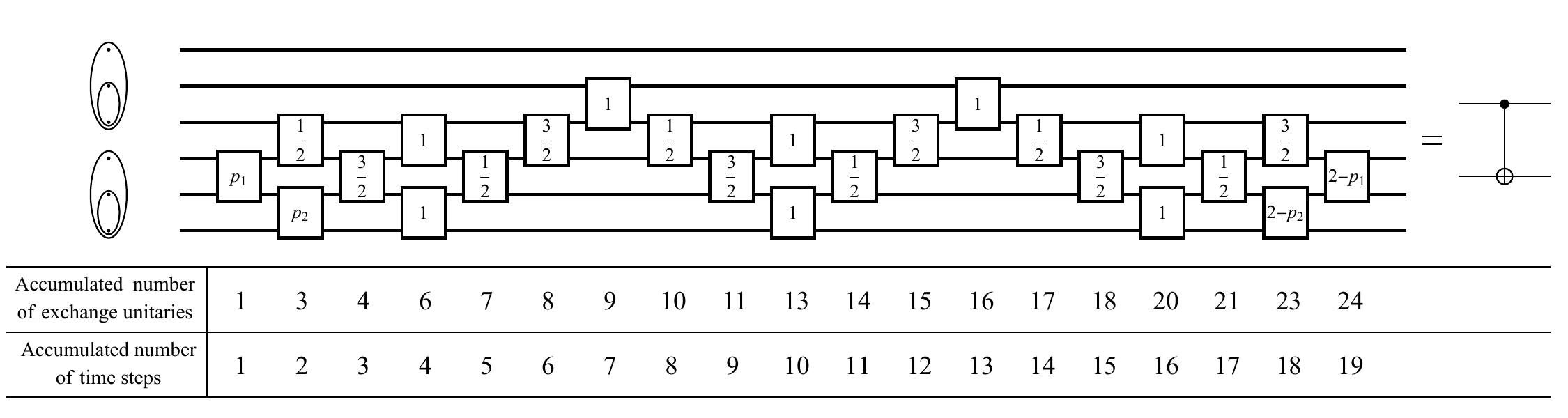}
		\caption{Quantum circuit illustrating the equivalent Fong-Wandzura sequence in the natural order (3,1,2,6,4,5) that implements a CNOT gate, with the accumulated number of exchange unitaries and number of time steps indicated. This sequence is the one used in this work.  Each block corresponds to an exchange unitary as defined in Eq.~\eqref{eq:fundamentalblock}, with the $p$ value indicated inside the block. Here $p_1=\arccos{\left(\frac{2\sqrt{3}}{3}-1\right)}/\pi$, $p_2=\arcsin{\left(\frac{2\sqrt{3}-1}{3}\right)}/\pi$.}
		\label{fig:312645FWsequence}
	\end{figure*}

Two-qubit gates are fundamental to quantum computation, with the CNOT gate being a particularly important and extensively studied example. DiVincenzo \emph{et al.} \cite{DiVincenzo2000,Makhlin2002} first proposed explicit pulse sequences for an encoded two-qubit gate that is locally equivalent to a CNOT gate. Their approach, consisting of 19 exchange unitaries over 13 time steps, was later shown to approximate an exact analytical solution \cite{Kawano2005}. Subsequently, Zeuch \emph{et al.} \cite{Zeuch.14} derived a fully analytical CNOT gate sequence for the entire subsystem, requiring approximately \blue{40 exchange unitaries}. Fong and Wandzura \cite{Bryan2011} introduced an improved analytic pulse sequence for the encoded CNOT gate, which achieves the same operation using only 22 exchange \blue{unitaries} in 13 time steps, as shown in Fig.~\ref{fig:312456FWsequence}. This represents the most efficient implementation of the CNOT gate in EO qubit systems to date. We emphasize again that
 the original Fong-Wandzura sequence is with natural order of (3,1,2,4,5,6). The one for natural order of (3,1,2,6,4,5),  implemented in this work, has a slightly different number of exchange unitaries which is shown in Fig.~\ref{fig:312645FWsequence}.

	\subsection{Triple EO qubits system}
	
We now consider a linear array of three EO qubits, labeled A, B, and C, arranged in the natural order (3,1,2,6,4,5,9,7,8) as shown in Fig.~\ref{fig:bulletrepresentation}(b).
The corresponding Hilbert space is spanned by $2^9 = 512$ basis states. However, the problem can be simplified by classifying the states using nine quantum numbers: $\{S_{\mathrm{tot}}^{(9)}, S_{z,\mathrm{tot}}^{(9)}, S_{A,B}, S_A, S_B, S_C, S_{A,1,2}, S_{B,1,2}, S_{C,1,2}\}$. Here, $S_{\mathrm{tot}}^{(9)}$ and $S_{z,\mathrm{tot}}^{(9)}$ denote the total spin and total $z$-component of spin of the full three-qubit system, with the superscript “$(9)$” indicating that these quantities are defined over all nine spins. $S_{A,B}$ is the total spin of qubits $A$ and $B$; $S_A$ ($S_B$, $S_C$) denotes the total spin of qubit $A$ ($B$, $C$); and $S_{A,1,2}$ ($S_{B,1,2}$, $S_{C,1,2}$) represents the total spin of particles 1 and 2 in \emph{natural order} within qubit $A$ ($B$, $C$). The eight computational basis states can be represented as follows in bullet notation:
	\begin{equation}
		\left|abc\right\rangle=\left(\bullet\left(\bullet\bullet\right)_a\right)_{1/2}\left(\bullet\left(\bullet\bullet\right)_b\right)_{1/2}\left(\bullet\left(\bullet\bullet\right)_c\right)_{1/2},
	\end{equation}
where $a,b,c$ are either $\mathbb{0}$ or $\mathbb{1}$ depending on the state of its corresponding qubit $A,B,C$ respectively.

While the complete basis set consists of 512 states, only a subset of these is relevant for our analysis, as the focus is on implementing three-qubit gates within the DFS. Since the DFS of a single EO qubit corresponds to the spin-$1/2$ subspace, applying three-qubit gates to the DFS requires targeting the matrix block associated with states where $S_A$, $S_B$, and $S_C$ are all $1/2$. These states are referred to as the computational bases. 
In the total angular momentum basis, 
the full $512\cross512$ matrix exhibits a highly block-diagonal structure. The key to reducing the basis size is to identify the smallest block matrix that encompasses all computational basis states, which is sufficient for implementing the desired quantum gates. Since the interaction between EO qubits consists solely of exchange couplings which commutes with the total angular momentum operator $S_{\mathrm{tot}}^{(9)}$, the matrix for a given value of $S_{\mathrm{tot}}^{(9)}$ naturally exhibits a block-diagonal structure. Among these blocks, only those corresponding to $S_{\mathrm{tot}}^{(9)}=1/2$ and $S_{\mathrm{tot}}^{(9)}=3/2$ can contain the states where $S_A$, $S_B$, and $S_C$ are all $1/2$. Consequently, the full $512\cross512$ matrix can be effectively reduced to the two relevant blocks associated with $S_{\mathrm{tot}}^{(9)}=1/2$ and $S_{\mathrm{tot}}^{(9)}=3/2$. Since the exchange coupling is rotationally invariant and commutes with $\boldsymbol{S}_{\mathrm{tot}}^{(9)}$, the Hamiltonian is further block-diagonal with respect to $S_{z,\mathrm{tot}}^{(9)}$.
This symmetry implies that it is sufficient to consider only one specific value of $S_{z,\mathrm{tot}}^{(9)}$ (in this work, we consistently select the maximum value). As a result, the two relevant blocks can be further reduced to a matrix constructed from only 90 basis states.

	\begin{table}[htb]
	\centering
	\renewcommand{\arraystretch}{2} 
	\setlength{\tabcolsep}{5pt} 
	\begin{tabular}{|c|c c c c|}
		\hline
		$S_{\mathrm{tot}}^{(9)}$ & $S_{A,B}$ & $S_C$ & \text{Number of States} & \text{Computational} \\ \hline
		$\frac{1}{2}$ & $0$ & $\frac{1}{2}$ & $5\cross2=10$ & Y \\ \hline
		$\frac{1}{2}$ & $1$ & $\frac{1}{2}$ & $9\cross2=18$ & Y \\ \hline
		$\frac{1}{2}$ & $1$ & $\frac{3}{2}$ & $9\cross1=9$ & N \\ \hline
		$\frac{1}{2}$ & $2$ & $\frac{3}{2}$ & $5\cross1=5$ & N \\ \hline
		$\frac{3}{2}$ & $1$ & $\frac{1}{2}$ & $9\cross2=18$ & Y \\ \hline
		$\frac{3}{2}$ & $2$ & $\frac{1}{2}$ & $5\cross2=10$ & N \\ \hline
		$\frac{3}{2}$ & $0$ & $\frac{3}{2}$ & $5\cross1=5$ & N \\ \hline
		$\frac{3}{2}$ & $1$ & $\frac{3}{2}$ & $9\cross1=9$ & N \\ \hline
		$\frac{3}{2}$ & $2$ & $\frac{3}{2}$ & $5\cross1=5$ & N \\ \hline
		$\frac{3}{2}$ & $3$ & $\frac{3}{2}$ & $1\cross1=1$ & N \\ \hline
	\end{tabular}
	\caption{Classification of the 90 basis states relevant for the triple-EO-qubit system, selected from the $S^{(9)}_{\mathrm{tot}}=1/2$ and $3/2$ subspaces.}
	\label{tab:necessary bases}
	\end{table}

	Table~\ref{tab:necessary bases} provides a detailed explanation of how the 90 basis states used in this work are systematically selected from the two relevant blocks corresponding to $S_{\mathrm{tot}}^{(9)}=1/2$ and $S_{\mathrm{tot}}^{(9)}=3/2$. The second and third columns list all possible combinations of $S_{A,B}$ and $S_C$ contributing to $S_{\mathrm{tot}}^{(9)}=1/2$ and $S_{\mathrm{tot}}^{(9)}=3/2$, respectively. The fourth column calculates the total number of bases by taking the product of the possible state counts in the second and third columns, thereby summarizing the selection process. The detail of these 90 bases are listed explicitly in the Supplemental Material.
	
	The last column of Table~\ref{tab:necessary bases} checks whether there exists a computational subspace in the  one spanned by the bases in the corresponding row. A computational subspace is composed of 8 computational bases. Under these bases, an arbitrary three-qubit gate can then be expressed as the following $90\times90$ matrix structure:
	\begin{equation}
		\left(
		\begin{array}{cccccc}
			U^{\text{comp}}_{8 \times 8} & 0 & 0 & 0 & 0 & 0 \\
			0 & U^{\text{leak}}_{2 \times 2} & 0 & U^{\text{leak}}_{2 \times 24} & 0 & 0 \\
			0 & 0 & U^{\text{comp}}_{8 \times 8} & 0 & 0 & 0 \\
			0 & U^{\text{leak}}_{24 \times 2} & 0 & U^{\text{leak}}_{24 \times 24} & 0 & 0 \\
			0 & 0 & 0 & 0 & U^{\text{comp}}_{8 \times 8} & 0 \\
			0 & 0 & 0 & 0 & 0 & U^{\text{leak}}_{40 \times 40}
		\end{array}
		\right).
	\end{equation}
	
	\begin{figure}[htb]
		\centering
		\includegraphics[width=\columnwidth]{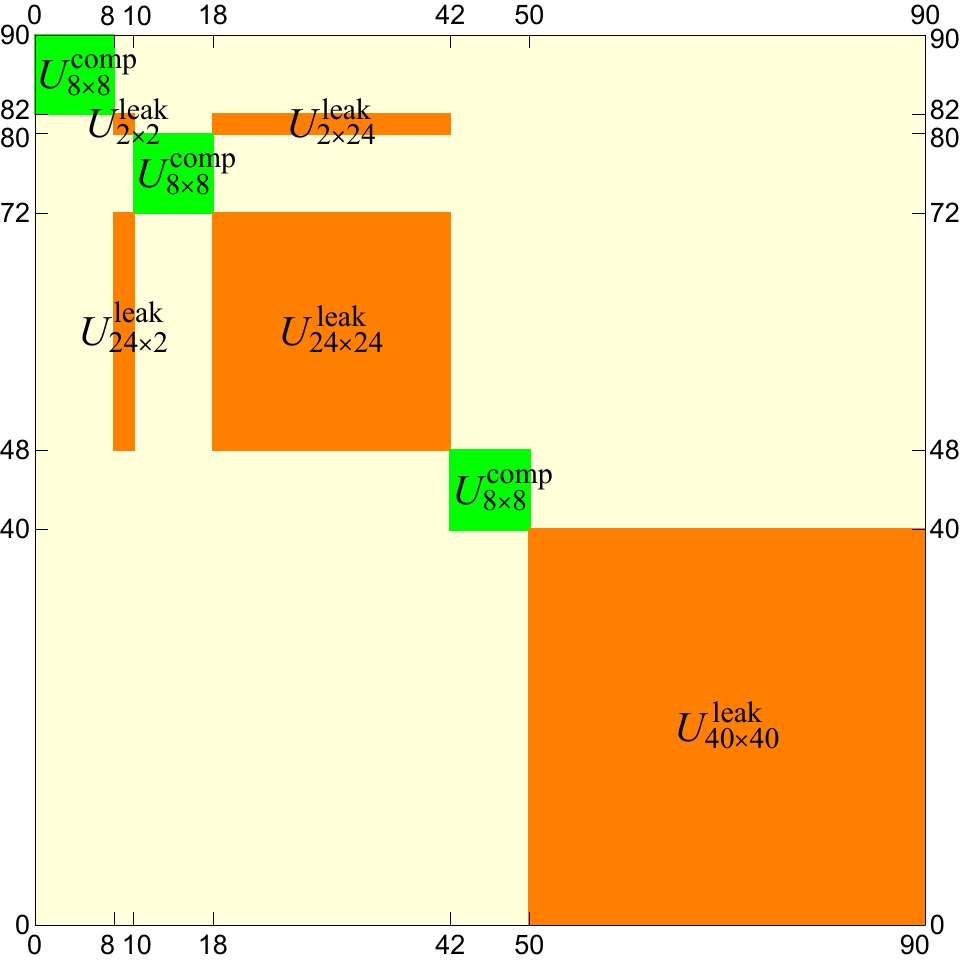}
		\caption{Block matrix structure of the reduced $90 \times 90$ Hamiltonian in the total angular momentum basis, showing computational (green) and leakage (orange) subspaces. Yellow regions are identically zero. Three identical $8 \times 8$ computational subspaces appear along the diagonal.}
		\label{fig:show9090matrix}
	\end{figure}
	
\begin{figure}[htb]
		\centering
		\includegraphics[width=\columnwidth]{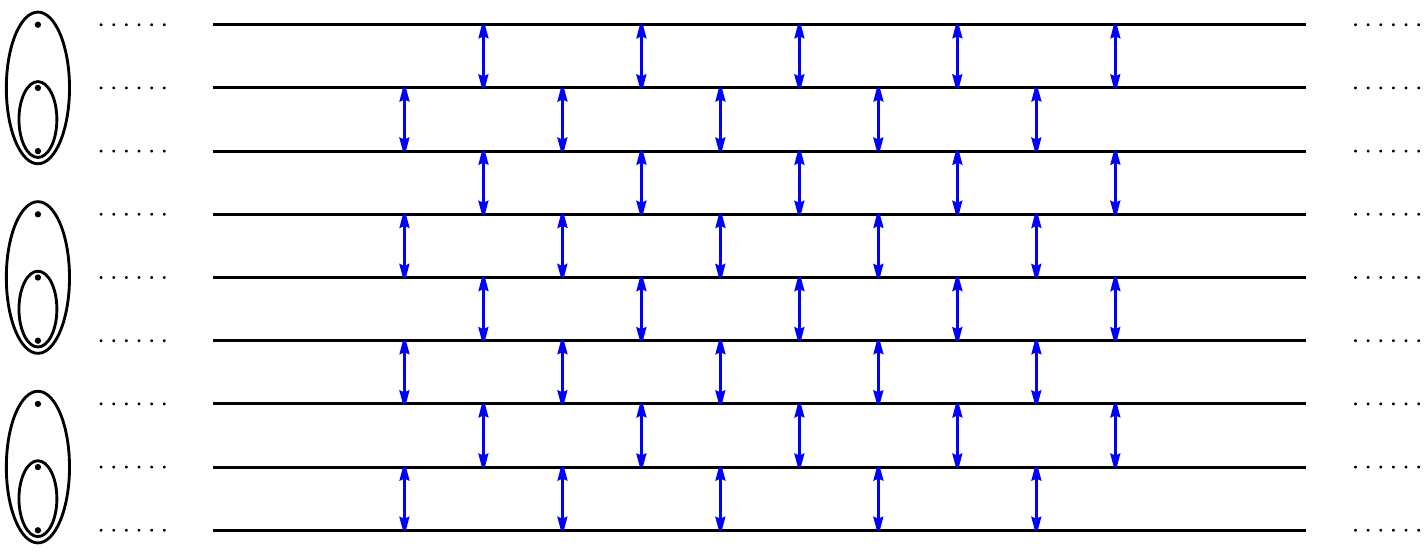}
		\caption{Schematic illustration of the compilation of three-qubit gates in a triple-EO-qubit system into exchange unitaries. The horizontal black lines represent individual spins and vertical blue arrows indicate controllable exchange unitaries.}
		\label{fig:controlstructure}
	\end{figure}

	Here the superscript ``comp'' denotes the computational subspace, while ``leak'' refers to subspaces involving leaked states. Fig.~\ref{fig:show9090matrix} depicts the block structure of this $90\cross90$ matrix. The regions highlighted in yellow correspond to matrix elements that are identically zero. The green blocks represent the subspaces spanned by the states involved in the implementation of the target three-qubit gate, while the orange blocks indicate the subspaces associated with leaked states. Due to permutation symmetry and block duplication under spin projection, three identical $8 \times 8$ computational subspaces appear along the diagonal. Therefore, any three-qubit gate that acts on the DFS is encoded in the upper-left-most block $U^{\text{comp}}_{8\times8}$. The main goal of this work is to find an efficient compilation, using elementary exchange unitaries to realize a Toffoli gate for this $U^{\text{comp}}_{8\times8}$, as shown in Fig.~\ref{fig:controlstructure}.

	\begin{figure*}[ht!]
		\centering
		\includegraphics[width=1.6\columnwidth]{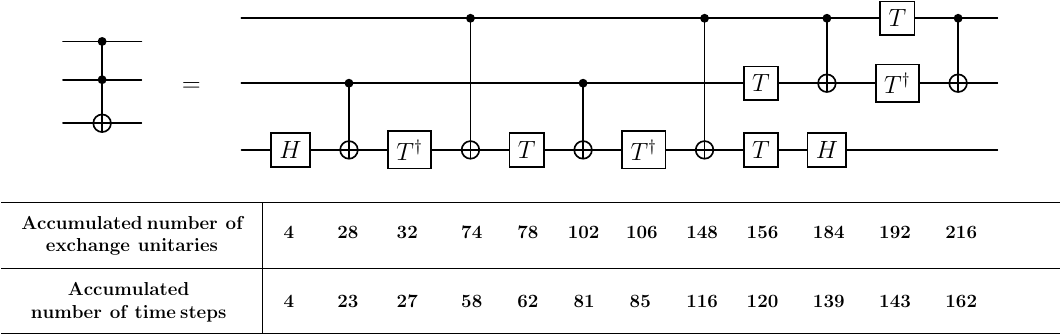}
		\caption{Quantum circuit showing direct decomposition of Toffoli gate into T, H and CNOT gates, namely the DIR sequence.}
		\label{fig:decomposition}
	\end{figure*}
	
	\section{Methods}\label{sec:methods}
	
	In this section, we discuss different approaches to decomposing a Toffoli gate into exchange unitaries. Section~\ref{subsec:directdecomp} introduces the DIR method, which, while conceptually straightforward, is highly inefficient --- requiring 216 exchange unitaries and 162 time steps. Numerical optimization methods, on the other hand, pose significant challenges due to the vast and complex parameter space and the intricate structure of the EO qubit Hamiltonian. Traditional algorithms often struggle to find global optima, frequently getting trapped in local minima or stalling on barren plateaus in the fidelity landscape, which leads to poor convergence or unacceptably low gate fidelities in practice. In Section~\ref{subsec:grape}, we examine the Gradient Ascent Pulse Engineering (GRAPE) method, followed by a discussion of Krotov’s method in Section~\ref{subsec:krotov}. Neither method proves satisfactory in solving our problem. To address these limitations, in Section~\ref{subsec:JK}, we present the JK algorithm, the key contribution of this work, and demonstrate that it offers a practical and high-fidelity solution for Toffoli gate design in EO architectures, overcoming these difficulties through a hybrid update rule that balances convergence stability and global search capability.

	\subsection{DIR}\label{subsec:directdecomp}

A straightforward yet ultimately impractical approach to implementing a Toffoli gate is via direct gate decomposition into single- and two-qubit gates, notably T, H and CNOT gates. However, this method typically yields excessively long pulse sequences, significantly undermining the efficiency of this fundamental operation. As illustrated in Fig.~\ref{fig:decomposition}, a typical decomposition consists of 2 Hadamard gates, 7 T (or T$^{\dagger}$) gates, 6 CNOT gates, and 4 SWAP gates \cite{Vivek2008}, resulting in a sequence that requires at least 216 exchange unitaries and 162 time steps. We refer to this as the ``DIR'' sequence for comparison with the JK algorithm results presented in Section~\ref{sec:result}. The presence of 4 SWAP gates in the decomposition arises from the connectivity constraints of the qubits. Specifically, among the 6 CNOT gates, four are between nearest-neighbor qubits and can be directly implemented using the equivalent FW sequence, as shown in Fig.~\ref{fig:312645FWsequence}. However, the remaining two CNOT gates act between non-nearest-neighbor qubits. To realize these operations, each non-nearest-neighbor CNOT gate requires two SWAP gates: one to bring the involved qubits adjacent to each other before applying the CNOT, and another to return them to their original positions afterward. Thus, a total of 4 SWAP gates are needed for these two non-nearest-neighbor CNOT gates. Such high overhead makes direct decomposition impractical in realistic quantum architectures, where Toffoli gates are frequently used and low-depth implementations are essential.
	
	\subsection{GRAPE method}\label{subsec:grape}
GRAPE is a widely adopted numerical optimization algorithm in quantum control, particularly effective for synthesizing high-fidelity unitary gates in a wide range of quantum systems. Its primary goal is to optimize a sequence of control pulses that guides a quantum system from a given initial state to a desired target state, or equivalently, to implement a specific unitary transformation. By iteratively maximizing a fidelity function using gradient-based techniques, GRAPE offers a robust and flexible framework for addressing a broad class of quantum control tasks.

The evolution of quantum systems is governed by the time-dependent Schr\"odinger equation:
	\begin{equation}
		\frac{d}{dt} U(t) = -i \mathcal{H}(t) U(t),
	\end{equation}
	where $U(t)$ is the time-evolution operator, and $\mathcal{H}(t)$ is the Hamiltonian of the system, which can typically be expressed as:
	\begin{equation}
		\mathcal{H}(t) = \mathcal{H}_0 + \sum_k u_k(t) \mathcal{H}_k.
	\end{equation}
	Here, $\mathcal{H}_0$ represents the drift Hamiltonian of the system, $\mathcal{H}_k$ are the control Hamiltonians, and $u_k(t)$ are the time-dependent control fields to be optimized.
	
	The fidelity of the control process is often defined in terms of the overlap between the achieved unitary $U(T)$ at the final time $T$ and the target unitary $U_{\text{target}}$. For unitary control, the fidelity function is typically expressed as:
	\begin{equation}
		\Phi = \frac{1}{d} \text{Re} \left( \text{Tr} \left[ U_{\text{target}}^\dagger U(T) \right] \right),
	\end{equation}
	where the dimension of the Hilbert space $d=24$.
	
	To optimize the fidelity $\Phi$, GRAPE calculates the gradient of $\Phi$ with respect to the control fields $u_k(t)$. The gradient can be efficiently computed using the chain rule and the propagators at intermediate time steps. Specifically, the gradient with respect to $u_k(t_j)$ (the control amplitude of the $k$-th field at time step $t_j$ is given by:
	\begin{equation}
		\begin{aligned}
			\frac{\partial\Phi}{\partial u_k(t_j)}
			&=\frac{2}{d} \mathrm{Re} \Bigg(\mathrm{Tr} \Big[ U_{\text{target}}^\dagger U(T) U(t_{N_T-1}) \cdots U(t_{j+1}) \\
			&\left( -i \frac{T}{N_T} \mathcal{H}_k \right) U(t_{j-1}) \cdots U(t_1) U(0) \Big]\Bigg)
		\end{aligned}
	\end{equation}
	where $N_T$ is the number of time steps, and the terms $U(0), U(t_1) \dots, U(t_{N_T-1}), U(T)$ represent the time-evolution operators at different steps.
	
	The optimization process in GRAPE involves iteratively updating the control fields $u_k(t)$ using a gradient-based optimization algorithm, such as gradient ascent or conjugate gradient. The update rule can be written as:
	\begin{equation}
		u_k^{(i+1)}(t_j) = u_k^{(i)}(t_j) + l_r \frac{\partial \Phi}{\partial u_k(t_j)},
	\end{equation}
	where $l_r$ is the learning rate, and $i$ denotes the iteration index.
	
GRAPE has been successfully employed across a wide range of quantum control tasks, including quantum gate synthesis, state preparation, and dynamical decoupling, owing to its computational efficiency and flexibility. It is considered a cornerstone in quantum optimal control. However, as will be demonstrated in the next subsection, GRAPE performs poorly in our context: it fails to achieve high fidelity even when optimizing sequences that are longer than those produced by our JK algorithm. This underscores the need for tailored optimization strategies that can effectively navigate the constrained and highly structured landscape of EO qubit systems.

	\begin{figure}[h]
		\centering
		\includegraphics[width=\columnwidth]{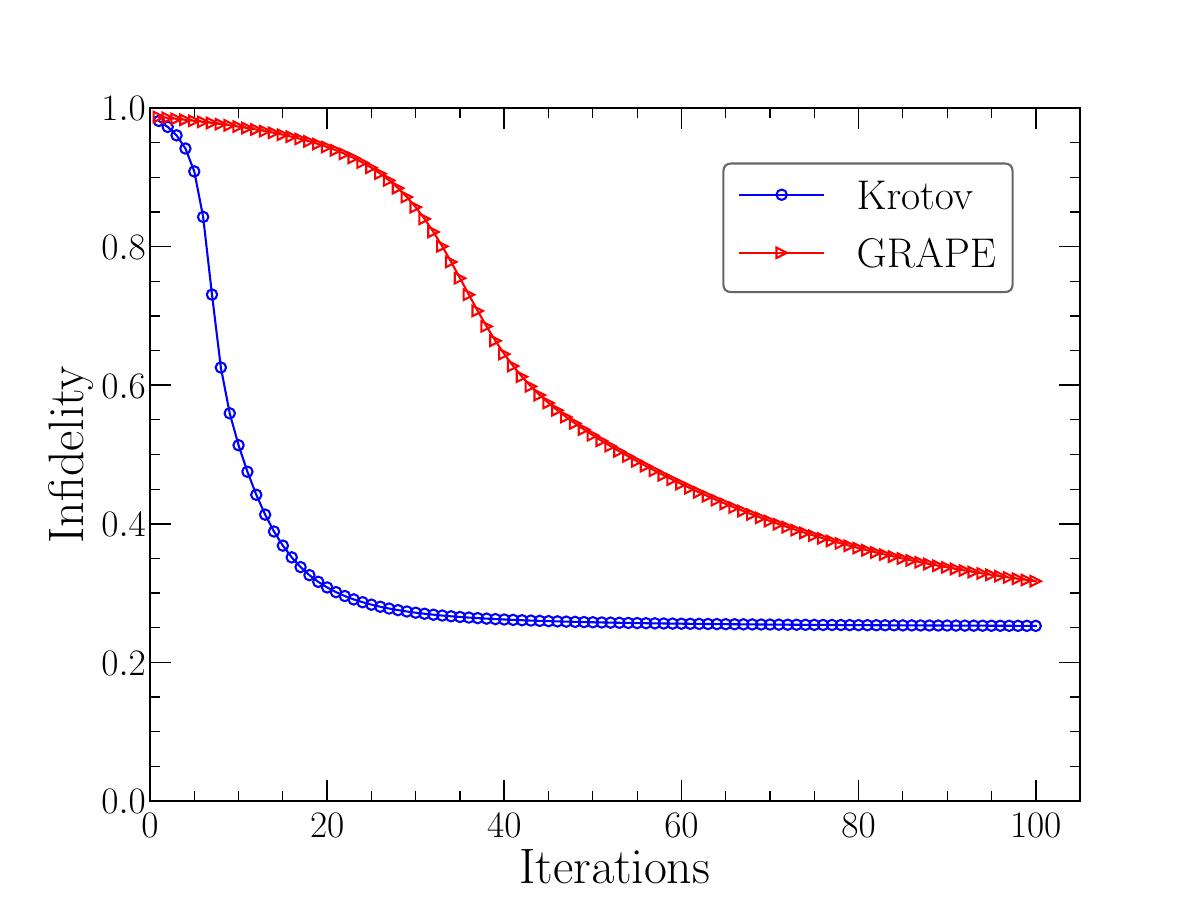}
		\caption{Fidelity convergence of GRAPE and Krotov’s method in the task of synthesizing a Toffoli gate using exchange unitaries. The optimization for both GRAPE and Krotov's method was carried out over 100 iterations with $N_T = 25$, which is too short to achieve lower infidelity. The learning rate for GRAPE was set to $l_r = 0.02$.}
		\label{fig:comparison}
	\end{figure}

	\subsection{Krotov's method}\label{subsec:krotov}

	Optimization algorithms can generally be categorized into gradient-free and gradient-based methods, depending on whether they rely solely on the evaluation of the cost functional or also incorporate its gradients. Among gradient-based approaches, Krotov’s method has gained widespread use in quantum optimal control applications \cite{Basilewitsch2020,Morzhin2019}. Its principal advantage lies in its suitability for problems involving continuous-time control sequences, a feature that aligns well with the sequential nature of our task \cite{Michael2019,Fernandes2023}. Unlike GRAPE, Krotov’s method ensures monotonic convergence of the fidelity under certain conditions, which makes it especially attractive for constrained control landscapes.
	
To minimize the total number of pulses required for implementing a gate, it is essential to design an efficient structure of control channels, thereby improving control-space utilization. Notably, the commutation relation $\left[\mathcal{H}_{i,j},\mathcal{H}_{k,l}\right] = 0$ holds when all indices $i,j,k,l$ are distinct. This observation enables the use of an interleaving strategy, in which the control Hamiltonian alternates between two non-overlapping sets of interactions, defined as:
	\begin{equation}
		\mathcal{H}(t) = 
		\begin{cases} 
			\mathcal{H}_\mathrm{odd}, & \text{for odd time steps,} \\
			\mathcal{H}_\mathrm{even}, & \text{for even time steps,}
		\end{cases}
	\end{equation}
	where $\mathcal{H}_\mathrm{even}=J_{1,2}\left(t\right)\boldsymbol{S}_1\cdot\boldsymbol{S}_2+J_{3,4}\left(t\right)\boldsymbol{S}_3\cdot\boldsymbol{S}_4+J_{5,6}\left(t\right)\boldsymbol{S}_5\cdot\boldsymbol{S}_6+J_{7,8}\left(t\right)\boldsymbol{S}_7\cdot\boldsymbol{S}_8$,  $\mathcal{H}_\mathrm{odd}=J_{2,3}\left(t\right)\boldsymbol{S}_2\cdot\boldsymbol{S}_3+J_{4,5}\left(t\right)\boldsymbol{S}_4\cdot\boldsymbol{S}_5+J_{6,7}\left(t\right)\boldsymbol{S}_6\cdot\boldsymbol{S}_7+J_{8,9}\left(t\right)\boldsymbol{S}_8\cdot\boldsymbol{S}_9$ [cf. Fig.~\ref{fig:controlstructure}]. These Hamiltonians are written in computational order. This interleaving exploits the commutativity of spatially disjoint exchange terms, enabling simultaneous updates in non-overlapping qubit pairs and effectively multiplying control channel utilization.

	The objective function of Krotov's method in this work is given by:
	\begin{equation}
		\begin{aligned}
			J_\mathrm{obj}&\left(\left\{\phi_k^{(i)}(t)\right\}, \left\{J_{l,l+1}^{(i)}(t)\right\}\right)=\\ &J_T\left(\left\{\phi_k^{(i)}(t)\right\}\right)
			+\sum_{l} \int_0^T g_a\left(J_{l,l+1}^{(i)}(t)\right)dt,
		\end{aligned}
	\end{equation}
	where $\left\{\phi_k^{(i)}(t)\right\}$ are the forward-propagated states initialized from $\left\{\phi_k\right\}$ under the control fields $\left\{J_{l,l+1}^{(i)}(t)\right\}$ at iteration $i$. For a three-qubit gate, the set $\left\{\phi_k\right\}$ corresponds to all eight logical basis states: $\left|\mathbb{000}\right\rangle$, $\left|\mathbb{001}\right\rangle$, $\left|\mathbb{010}\right\rangle$, $\left|\mathbb{011}\right\rangle$, $\left|\mathbb{100}\right\rangle$, $\left|\mathbb{101}\right\rangle$, $\left|\mathbb{110}\right\rangle$, and $\left|\mathbb{111}\right\rangle$.
	
	The first part of the objective function is a final time related term $J_T$, which plays major role in $J_\mathrm{obj}$. A straight-forward choice of $J_T$ for our task is the total state infidelity:
	\begin{equation}
		J_{T,\mathrm{re}} = 1 - \frac{1}{N} \text{Re} \left[\sum_{k=1}^{N} \tau_k \right],
	\end{equation}
	with $\tau_k = \left\langle \phi_k^{\text{target}} \middle| \phi_k(T) \right\rangle$, $\left| \phi_k^{\text{target}} \right\rangle = U_\mathrm{gate} \left| \phi_k \right\rangle$, and $N$ being the dimension of the logical subspace.
	
	The second part of $J_\mathrm{obj}$ is the running cost on the control field. A commonly used expression of $g_a$ is:
	\begin{equation}
		g_a\left(J_{l,l+1}^{(i)}(t)\right)=\frac{\lambda_{l}}{S_l(t)}\left(J_{l,l+1}^{(i)}(t)-J_{l,l+1}^{(i-1)}(t)\right)^2,
	\end{equation}
	where the inverse step width $\lambda_{l}>0$ and the update shape function $S_l(t)$ can be any user-defined function whose values range between 0 and 1.
	
	After setting the initial guess $J_{l,l+1}^{(0)}(t)$, the optimized field $J_{l,l+1}^{(i)}(t)$ in iteration $i$ is updated as follows:
	\begin{equation}
		J_{l,l+1}^{(i)}(t) = J_{l,l+1}^{(i-1)}(t) + \Delta J_{l,l+1}^{(i)}(t),
	\end{equation}
	where
	\begin{equation}
	\begin{split}
		&\Delta J_{l,l+1}^{(i)}(t) = \\ 
		&\frac{S_l(t)}{\lambda_{l}} \text{Im} \left[ \sum_{k=1}^{N} \left\langle \chi_k^{i-1}(t) \middle| \left( \frac{\partial \hat{\mathcal{H}}}{\partial J_{l,l+1}} \right)_{(i)} \middle| \phi_k^{(i)}(t) \right\rangle \right],
	\end{split}\end{equation}
	with the forward propagation equation of motion for $\left| \phi_k^{(i)}(t) \right\rangle$ given by:
	\begin{equation}
		\frac{\partial}{\partial t} \left| \phi_k^{(i)}(t) \right\rangle = -\frac{i}{\hbar} \hat{\mathcal{H}}(t) \left| \phi_k^{(i)}(t) \right\rangle.
	\end{equation}
	The co-states $\left| \chi_k^{i-1}(t) \right\rangle$ are propagated backward in time under the guess controls of iteration $i-1$, satisfying:
	\begin{equation}
		\frac{\partial}{\partial t} \lvert \chi_k^{i-1}(t) \rangle = -\frac{i}{\hbar} \hat{\mathcal{H}}^{(i-1) \dagger} \lvert \chi_k^{i-1}(t) \rangle,
	\end{equation}
	with the boundary condition:
	\begin{equation}
		\lvert \chi_k^{i-1}(T) \rangle = - \left. \frac{\partial J_T}{\partial \langle \phi_k(T) \rangle} \right|_{(i-1)}.
	\end{equation}
	For our specific $J_T=J_{T,\mathrm{re}}$, this boundary condition gives:
	\begin{equation}
		\lvert \chi_k(T) \rangle = \frac{1}{2N} \left| \phi_k^{\text{target}} \right\rangle,
	\end{equation}
	which is independent of $\phi_k(T)$. The pseudocode of this procedure is shown as Algorithm~\ref{alg:krotov}.
	\begin{algorithm}[H]
		\caption{Krotov's Method}
		\begin{algorithmic}[1]
			\State set initial $\{J_{l,l+1}^{(0)}(t)\}$, $\{\phi_k\}$, $\{S_{l,n}\}$, $\{\lambda_l\}$, forward propagator $U$, backward propagator $U^{\dagger}$, infidelity threshold $\epsilon=10^{-8}$
			\State $i \gets 0$
			\State allocate forward storage array $\Phi_0[1 \ldots N(=24), 0 \ldots N_T]$
			\For{$k \gets 1, \ldots, N$}
				\State $\Phi_0[k,0] \gets \phi_k^{(0)}(t_0) \gets \phi_k$
				\For{$n \gets 0, \ldots, N_T$}
					\State $U(t_n) \gets H(t_n) \gets J_{l,l+1}^{(0)}(t_n)$
					\State $\Phi_0[k,n+1] \gets \phi_k^{(0)}(t_{n+1}) \gets U(\phi_k^{(0)}(t_n))$
				\EndFor
			\EndFor
			\State $J_{T,\mathrm{re}} \gets \Phi_0[k,N_T]$
			\While{not converged}
				\State allocate backward storage array $X[1 \ldots N, 0 \ldots N_T]$
				\For{$k \gets 1, \ldots, N$}
					\State $X[k,N_T] \gets \chi^{(i)}_k(T) \gets \phi_k^{\text{target}}/(2N)$
					\For{$n \gets N_T, \ldots, 0$}
						\State $U^{\dagger}(t_n) \gets H^{\dagger}(t_n) \gets J_{l,l+1}^{(i)}(t_n)$
						\State $X[k,n-1] \gets \chi^{(i)}_k(t_{n-1}) \gets U^{\dagger}(\chi^{(i)}_k(t_n))$
					\EndFor
				\EndFor
				\State allocate forward storage array $\Phi_1[1 \ldots N, 0 \ldots N_T]$
				\State $\forall k : \Phi_1[k,0] \gets \phi^{(i+1)}_k(t_0) \gets \phi_k$
				\For{$n \gets 0, \ldots, N_T$}
					\State $\forall k : \chi_k^{(i)}(t_n) = X[k,n]$
					
					\State $\forall l : \Delta J_{l,l+1}^{(i)}(t_n) \gets$
					\Statex $\qquad\qquad\qquad \frac{S_{l,n}}{\lambda_l} \operatorname{Im} \sum_k \langle \chi_k^{(i)}(t_n) |\frac{\partial H}{\partial J_{l,l+1}}| \phi^{(i+1)}_k(t_n) \rangle$
					\State $\forall l : J_{l,l+1}^{(i+1)}(t_n) \gets J_{l,l+1}^{(i)}(t_n) + \Delta J_{l,l+1}^{(i)}(t_n)$
					\State \State $U(t_n) \gets H(t_n) \gets J_{l,l+1}^{(i+1)}(t_n)$
					\State $\forall k : \Phi_1[k,n+1] \gets \phi_k^{(i+1)}(t_{n+1}) \gets U(\phi_k^{(i+1)}(t_n))$
				\EndFor
				\State $\Phi_0 \gets \Phi_1$
				\State $J_{T,\mathrm{re}} \gets \Phi_1[k,N_T]$
				\If{$J_{T,\mathrm{re}}<\epsilon$}
					break
				\EndIf
				\State $i \gets i+1$
			\EndWhile
			\State $\{J_{l,l+1}^{\text{opt}}(t)\} \gets \{J_{l,l+1}^{(i)}(t)\}$
		\end{algorithmic}\label{alg:krotov}
	\end{algorithm}
	
Fig.~\ref{fig:comparison} demonstrates that when the total number of time steps $N_T$ is restricted to a relatively small value, both GRAPE and Krotov’s method perform poorly in the Toffoli gate search task and fail to achieve low infidelity. Here, we set $N_T = 25$ for both algorithms. Because each time step contains four exchange unitaries (cf.~Fig.~\ref{fig:controlstructure}), this leads to a 100-pulse sequence---already longer than the 92-pulse sequence found by our JK algorithm. However, as illustrated in Fig.~\ref{fig:comparison}, after 100 iterations, the gate infidelity for GRAPE only approaches around 0.3, while Krotov’s method achieves better results but still plateaus at approximately 0.22. Both results are far from the infidelity achieved by the 92-pulse sequence found by our JK algorithm, which indicates that a single gradient-based algorithm alone is insufficient for this task under constrained $N_T$. In contrast, our JK algorithm circumvents this limitation by first performing gradient-based optimization with a much larger $N_T$ (e.g., $N_T = 55$, resulting in a 220-pulse sequence), and then pruning most of the unnecessary unitaries. Without the restriction of small $N_T$, it is always possible to use gradient methods (mainly Krotov’s method in this work) to optimize from multiple random initializations. If, for a large number of random initializations, it remains difficult to reduce the infidelity below the required threshold of $10^{-8}$, $N_T$ can be further increased until solutions meeting the threshold can be reliably found with only a small number of random initializations.

	
	\subsection{Jenga-Krotov algorithm}\label{subsec:JK}

Krotov’s method, discussed in the previous section, has proven valuable for quantum‐gate optimization. Yet our preliminary tests reveal that, when the final-gate infidelity is used directly as the cost functional, the original algorithm exhibits serious shortcomings for our problem. Because Krotov’s update scheme is strictly local in parameter space, its performance is highly sensitive to the initial guess and it frequently becomes trapped in poor local minima, leaving the residual infidelity unacceptably high for practical use.

\begin{figure*}[ht]
		\centering
		\includegraphics[width=2\columnwidth]{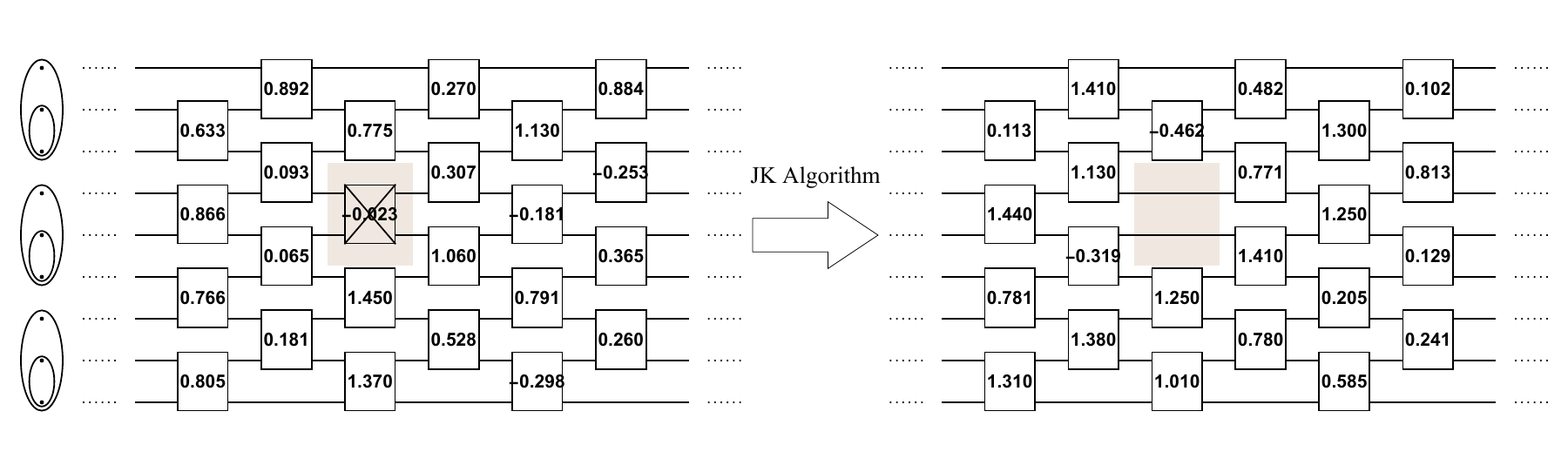}
		\caption{Illustration of the JK algorithm. The left panel shows an initial dense pulse sequence, where one exchange unitary (highlighted by the shaded box) is randomly selected and removed. The sequence is then re-optimized, allowing the durations of the remaining \blue{exchange unitaries} (indicated by the numbers in the boxes) to adjust. The right panel shows the resulting optimized sequence, now with one fewer exchange unitary, but maintaining the same level of gate fidelity. This iterative pruning-and-refinement process is key to the JK algorithm's ability to yield compact, high-fidelity pulse sequences.}
		\label{fig:brick transformation}
	\end{figure*}

To overcome these limitations we introduce the JK algorithm, which retains the strengths of Krotov’s framework while adding two complementary strategies that promote global convergence.
\begin{itemize}
\item \emph{Parameter-space expansion.} We first lengthen the initial pulse sequence, giving the optimizer a higher-dimensional search space in which to escape shallow minima and locate a configuration whose final infidelity falls below a chosen threshold.  One practical approach is to 
initialize with a randomly generated, sufficiently long pulse train
and use the Krotov algorithm to iteratively adjust it toward the target unitary. This step encourages exploration of wider neighborhoods in the control landscape, avoiding barren plateaus.
\item \emph{Systematic pulse reduction.} Once a high-fidelity pulse sequence is obtained, we initiate a pruning process: individual exchange unitaries (treated as a `brick') are tentatively removed, one at a time, and the resulting sequence is re-optimized. During this re-optimization, the locations of all remaining exchange unitaries are kept fixed, but the durations of the corresponding \blue{exchange unitaries} are allowed to vary. An exchange unitary is eliminated if its removal increases the final gate infidelity by less than a predetermined threshold. By iterating this “brick-removal” process (as illustrated in Fig.~\ref{fig:brick transformation}), we obtain a significantly more compact pulse sequence that retains the fidelity of its denser predecessor.
\end{itemize}

The name Jenga-Krotov reflects the core strategy of the algorithm: much like the game Jenga, which begins with a fully constructed tower and gradually removes blocks without compromising its stability, the JK algorithm starts with a dense, high-fidelity pulse sequence and systematically prunes away redundant exchange unitaries while maintaining the target performance.

This two-stage approach, i.e. first expanding the search space to escape local minima, then reducing complexity through guided pruning, not only mitigates the limitations of local gradient optimization but also significantly lowers computational overhead. Using JK, we are able to generate a Toffoli sequence that meets practical fidelity thresholds with substantially fewer \blue{exchange unitaries} than those produced by conventional Krotov or GRAPE methods.

\blue{To address the issue of sensitivity to initialization in the Krotov method, we systematically investigated different parameter settings, including step size and other hyperparameters, and explored adaptive tuning strategies. Our results show that while such adjustments can marginally improve performance, they do not fundamentally resolve the sensitivity inherent to the optimization landscape. Notably, we found that expanding the parameter space by increasing the sequence length $N_T$ is a much more effective approach. By dynamically adjusting $N_T$ based on preliminary optimization results, the Krotov method is more likely to reach a global optimum from various random initializations, as evidenced by achieving infidelity below a threshold. This reflects the existence of multiple global optima corresponding to different pulse sequences that can implement the same quantum gate.}

\blue{Our strategy involves incrementally increasing $N_T$ and conducting a pre-optimization sampling, where we run a limited number of Krotov iterations from several random initializations. If most runs fail to reach the desired infidelity, it signals the need for a larger $N_T$. This procedure is computationally efficient and enables rapid identification of suitable sequence lengths.}

\blue{However, simply increasing $N_T$ often leads to unnecessarily long and inefficient gate sequences. To address this, our JK algorithm incorporates a post-optimization pruning step, which refines the raw sequence into a more compact and efficient form. The combination of adaptive expansion of the parameter space and subsequent pruning is central to the robustness and efficiency of our approach, distinguishing the JK algorithm from straightforward modifications of the Krotov method.}

The complete procedure is summarized in Algorithm~\ref{algorithm:JK}.

	\begin{figure}[htb]
		\centering
		\includegraphics[width=\columnwidth]{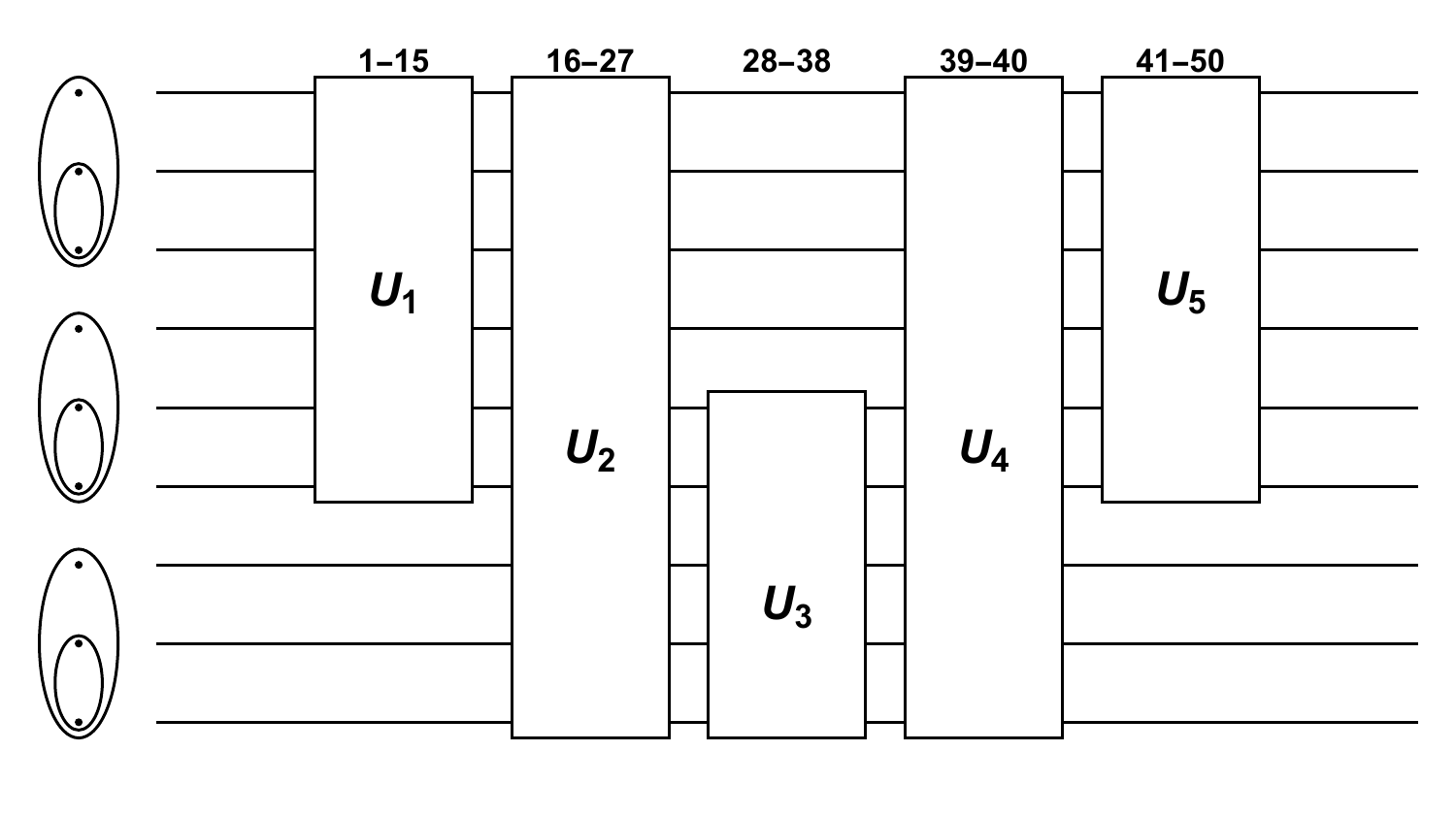}
		\caption{Illustration of the final optimized pulse sequence for the Toffoli gate obtained by the JK algorithm in five sections of unitaries.}
		\label{fig:pulse sequence structure}
	\end{figure}
	
	\begin{algorithm}[H]
		\caption{Jenga-Krotov Algorithm}
		\begin{algorithmic}[1]
			\State \textbf{Initialize:} Set a long initial pulse sequence of length $L$ and an infidelity threshold $\epsilon$.
			\State Use Krotov's method to find the optimal $\{J_{l,l+1}^L(t)\}$ such that the infidelity is less than $\epsilon$. If this fails, return to step 1 with a larger $L$.
			\State Allocate a $4 \times L$ matrix $M$ with all elements set to $-1$, indicating unexplored entries. At this stage, the number of $-1$ in $M$ is $4L$, which equals the number of exchange unitaries $N_{\text{ex}}$.
			\While{there exists $e = -1$ in $M$}
				\State Randomly select a position $(i, j)$ such that $M_{i, j} = -1$.
				\State $(i, j)$ refers to the $i$th control at time step $j$.
				\State Set $J_{i,i+1}^L(t_j) = 0$.
				\State Use Krotov's method to find a new optimum $\{J_{l,l+1}^{\prime L}(t)\}$.
				\If{$J_{T, \mathrm{re}}(\{J_{l,l+1}^{\prime L}(t)\}) < \epsilon$}
					\State $\{J_{l,l+1}^L(t)\} \gets \{J_{l,l+1}^{\prime L}(t)\}$
					\State $M_{i, j} \gets 0$
					\State $N_{\text{ex}} \gets N_{\text{ex}} - 1$
				\Else
					\State $M_{i, j} \gets 1$
				\EndIf
			\EndWhile
			\State $N^{\text{opt}}_{\text{ex}} \gets N_{\text{ex}}$
			\For{$j = 1$ \textbf{to} $L$}
				\If{$M_{i, j} = 0$ for all $i$}
					\State $L \gets L - 1$
				\EndIf
			\EndFor
			\State $L^{\mathrm{opt}} \gets L$
			\State $\{J_{l,l+1}^{\text{opt}}(t)\} \gets \{J_{l,l+1}^L(t)\}$
		\end{algorithmic}
		\label{algorithm:JK}
	\end{algorithm}
	
	\section{Results}\label{sec:result}
	\subsection{JK-based efficient pulse optimization for Toffoli gate}
	
	In this section, we present our results on efficiently decomposing the Toffoli gate using the JK algorithm. We begin by generating random initial control pulses that fully populate a dense, over-complete control structure—activating all allowable exchange couplings across both spatial and temporal channels, regardless of whether they ultimately contribute to the target gate. This initial sequence comprises 192 exchange unitaries over 55 time steps. Krotov’s method is then applied to optimize the pulse parameters, driving the system toward a Toffoli operation; however, it does not alter the sequence length or structure. The resulting high-fidelity but uncompressed sequence is detailed in the Supplemental Material. Although this uncompressed sequence successfully realizes the desired gate, its length and complexity render it impractical for direct use in scalable architectures.

	To address this, we applied the JK algorithm to systematically remove pulses and further optimize the pulse parameters. \blue{All optimizations were performed on a workstation equipped with an Intel Core Ultra 7 155H processor (1.40 GHz, 16 cores) and 32 GB RAM, with a typical total computational time of about 20 hours.} This process resulted in a significantly refined sequence consisting of 92 pulses distributed across 50 time steps, as detailed in Appendix~\ref{appx:sequence}. The optimized sequence is visually represented in Fig.~\ref{fig:pulse sequence structure}, where it is divided into five sections based on their active qubit subsets and functional roles:
	
	$U_1$: A unitary acting on qubits A and B, spanning time steps 1 to 15, consisting of 24 pulses.
	
	$U_2$: A unitary acting on all three qubits, spanning time steps 16 to 27, consisting of 29 pulses.
	
	$U_3$: A unitary acting on qubits B and C, spanning time steps 28 to 38, consisting of 18 pulses.
	
	$U_4$: A unitary acting on all three qubits, spanning time steps 39 to 40, consisting of 6 pulses.
	
	$U_5$: A unitary acting on qubits A and B, spanning time steps 41 to 50, consisting of 15 pulses.\\
	
	The final unitary operation implemented by this sequence is expressed as:
	\begin{equation}
		U_{\text{final}} = U_5 U_4 U_3 U_2 U_1,
	\end{equation}
	where each segment $U_i$ is defined as the product of individual unitary operations over its respective time steps:
	\begin{equation}
		\begin{aligned}
			&U_1 = \prod_{k=15}^{1} U^{(k)}, \quad U_2 = \prod_{k=27}^{16} U^{(k)}, \quad U_3 = \prod_{k=38}^{28} U^{(k)}, \\
			&U_4 = \prod_{k=40}^{39} U^{(k)}, \quad U_5 = \prod_{k=50}^{41} U^{(k)},
		\end{aligned}
	\end{equation}
	and the individual unitary operations $U^{(k)}$ are defined as:
	\begin{equation}
		U^{(k)} = 
		\begin{cases} 
			\begin{split}
			e^{-i\pi\left(p_{2,3}^{(k)}\mathcal{H}_{2,3}^\mathrm{ex}+p_{4,5}^{(k)}\mathcal{H}_{4,5}^\mathrm{ex}+p_{6,7}^{(k)}\mathcal{H}_{6,7}^\mathrm{ex}+p_{8,9}^{(k)}\mathcal{H}_{8,9}^\mathrm{ex}\right)} & \\ 
			(\text{for}& \text{\ odd\ } k),
			\end{split} \\
			\begin{split}
			e^{-i\pi\left(p_{1,2}^{(k)}\mathcal{H}_{1,2}^\mathrm{ex}+p_{3,4}^{(k)}\mathcal{H}_{3,4}^\mathrm{ex}+p_{5,6}^{(k)}\mathcal{H}_{5,6}^\mathrm{ex}+p_{7,8}^{(k)}\mathcal{H}_{7,8}^\mathrm{ex}\right)} & \\ 
			(\text{for}& \text{\ even\ } k),
			\end{split}
		\end{cases}\label{eq:Ukstep}
	\end{equation}
	with
	\begin{equation}
		\mathcal{H}_{i,j}^\mathrm{ex}=\frac{1}{4}\boldsymbol{\sigma}_i\cdot\boldsymbol{\sigma}_j-\frac{1}{4},\left(i,j=1,2,3\right).
	\end{equation}
	Here the labels are in the computational order. The 92-pulse sequence obtained via the JK algorithm offers significant advantages over the DIR sequence. As summarized in Table~\ref{tab:sequence comparison}, the JK approach yields a markedly shorter and more efficient pulse sequence for realizing the Toffoli gate, substantially reducing both the total number of exchange unitaries and time steps while preserving high fidelity. \blue{In our optimization, the maximum and minimum values of the exchange coupling $J$ are selected to match the experimentally accessible range for exchange-only qubits. Modern semiconductor quantum dot devices allow $J$ to be tuned from below 1~MHz up to approximately 10~GHz \cite{Qiao2020}. Our simulations ensure that all $J$ values remain within these practical limits, with the minimum $J$ set to a small finite value for numerical stability. For example, with $J_0 = 100$ MHz, each time step in the JK sequence is approximately 31.4 ns, and a 50-step sequence results in a total gate time of about 1.57~$\mu$s. This duration is well below the typical decoherence time ($T_2^*$) of exchange-only qubits, which often exceeds 2~$\mu$s with advanced control techniques. Thus, all pulse amplitudes and gate durations used in our optimization are compatible with current experimental capabilities.} These improvements demonstrate the JK algorithm’s effectiveness in addressing the inefficiencies of direct decomposition methods and affirm its potential for broader applications in high-performance quantum control in EO qubit systems.
	
	\blue{To further illustrate the flexibility and efficiency of the JK algorithm, we have extended our study to multi-qubit gates and complex quantum circuits, such as the quantum Fourier transform (QFT) circuit and the Fredkin gate. Direct decomposition methods for these circuits typically result in long pulse sequences with a large number of time steps and exchange unitaries---for example, the QFT circuit requires 180 time steps and 237 exchange unitaries, while the Fredkin gate requires 200 time steps and 276 exchange unitaries. In contrast, the JK algorithm can directly optimize the entire circuit, yielding significantly shorter and more efficient pulse sequences: only 80 time steps and 202 exchange unitaries for the QFT circuit, and 104 time steps and 172 exchange unitaries for the Fredkin gate. This demonstrates that the JK algorithm can efficiently optimize pulse sequences for complex quantum algorithms and multi-qubit gates, not just individual cases. Detailed results and comparisons are provided in the Supplemental Material and our GitHub repository~\cite{data}.}

\begin{table}[tb]
		\centering
		\renewcommand{\arraystretch}{2} 
		\setlength{\tabcolsep}{7pt} 
		\begin{tabular}{|c|c|c|}
			\hline
			Property & DIR sequence & JK sequence \\ \hline
			Total time steps & 162 & 50 \\ \hline
			\begin{tabular}{@{}c@{}}  Total exchange unitaries \end{tabular}   & 216 & 92 \\ \hline
			maximum $J/J_0$ & 1 & 1.342 \\ \hline
			minimum $J/J_0$ & -1/2 & -1.592 \\ \hline
		\end{tabular}
		\caption{Comparison of key properties between DIR sequence and the JK-optimized sequence.}
		\label{tab:sequence comparison}
	\end{table}
	
	\subsection{Performance under noisy environment}
	
	In this section, we evaluate the robustness of the optimized 92-pulse Toffoli gate sequence under realistic noise conditions. Our noise model incorporates both control-dependent and quasi-static charge noise~\cite{Hu.06, Kestner2013, Reed.16, Keith.22}, which are characterized by fluctuations in the exchange coupling strength:
	\begin{equation}
		p_{i,i+1} \rightarrow (1+\alpha)p_{i,i+1},
	\end{equation}
	where $\alpha$ represents a quasi-static noise parameter that is assumed to be identical for all $p_{i,i+1}$ in a given realization. we also consider crosstalk effects between neighboring spins, modeled by the transformation:
	\begin{equation}
		p_{i,i+1}\mathcal{H}_{i,i+1}^\mathrm{ex}\rightarrow p_{i,i+1}\left[\mathcal{H}_{i,i+1}^\mathrm{ex}+\beta\left(\mathcal{H}_{i-1,i}^\mathrm{ex}+\mathcal{H}_{i+1,i+2}^\mathrm{ex}\right)\right],
	\end{equation}
	where $\beta$ represents the strength of crosstalk which is also assumed to be identical for all $p_{i,i+1}$ in a given realization. 
	
	In our analysis, we model $\alpha$ and $\beta$ as random variables drawn from normal distribution $N(\bar{\alpha},0.1\bar{\alpha})$ and $N(\bar{\beta},0.1\bar{\beta})$ respectively, where the mean value $\bar{\alpha}$ and $\bar{\beta}$ are varied from $10^{-8}$ to $10^{-1}$.
	
	\blue{Typical magnitudes of quasi-static charge noise and crosstalk in exchange-only qubit systems have been well characterized in recent studies. Quasi-static charge noise is usually modeled as Gaussian fluctuations in the exchange coupling, with standard deviation $\sigma_J$ commonly in the range $0.01J$ to $0.1J$ \cite{Throckmorton2022}. Current devices can achieve $\sigma_J \approx 0.01J$, though higher values may occur. This noise significantly affects gate fidelity, especially for longer sequences, making optimized pulse design essential. Crosstalk, arising from unwanted exchange interactions between neighboring qubits, is typically below $0.5\%$ of the maximal exchange coupling and can be further reduced to below $0.1\%$ with advanced compensation and pulse engineering \cite{Madzik2025}. At these levels, the impact on gate fidelity is negligible. Continued improvements in device design and control are expected to further suppress both charge noise and crosstalk.}
	
	To quantify the effect of noise, we compute the gate fidelity using the standard formula:
	\begin{equation}
		F=\frac{d+\left|\Tr\left(U_{\text{ideal}}^{\dagger}U_{\text{actual}}\right)\right|^2}{d\left(d+1\right)},
	\end{equation}
	where $U_{\text{ideal}}$ denotes the target Toffoli operation, $U_{\text{actual}}$ is the noisy implementation of the JK-sequence and $d=24$ is the total dimension of the three $8 \times 8$ blocks.
	
\begin{figure}[htb]
		\centering
		(a)\includegraphics[width=0.5\textwidth]{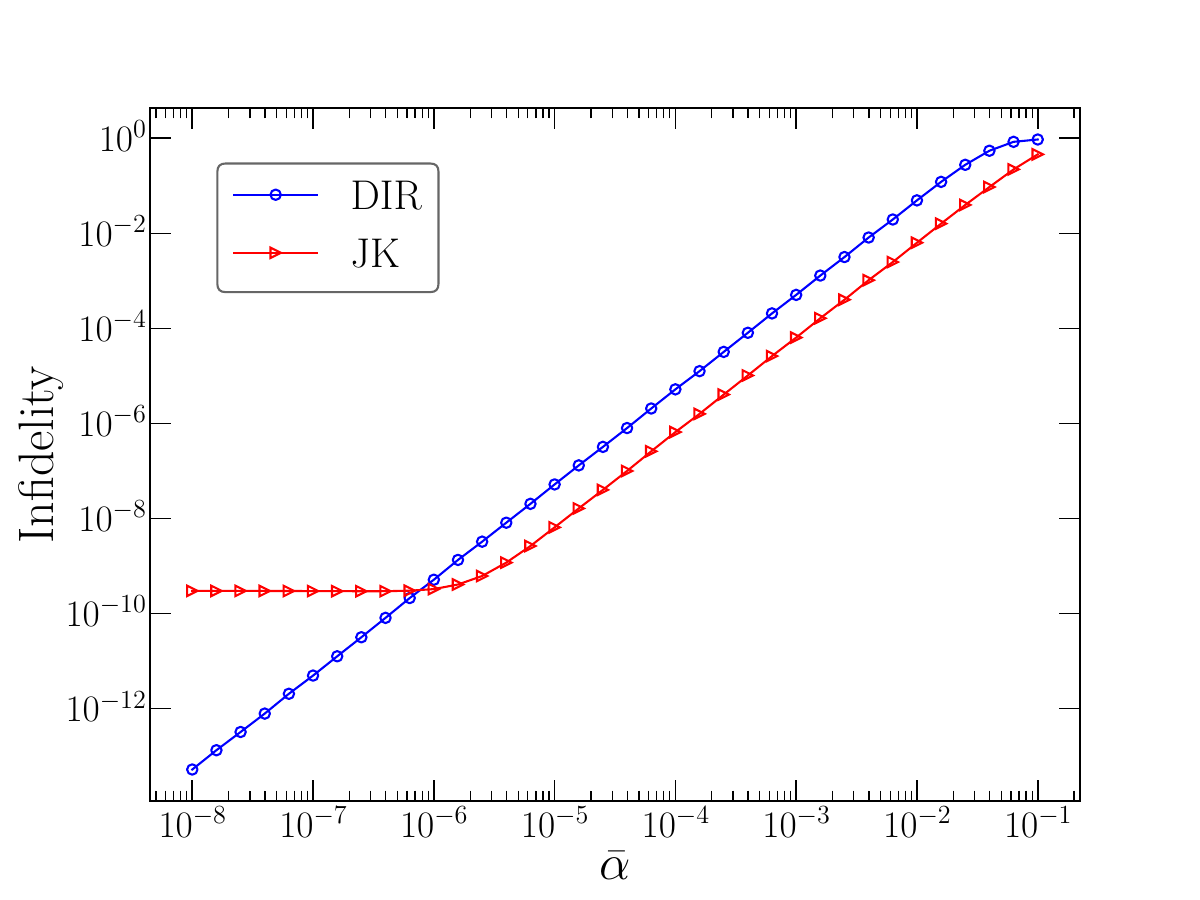}
		(b)\includegraphics[width=0.5\textwidth]{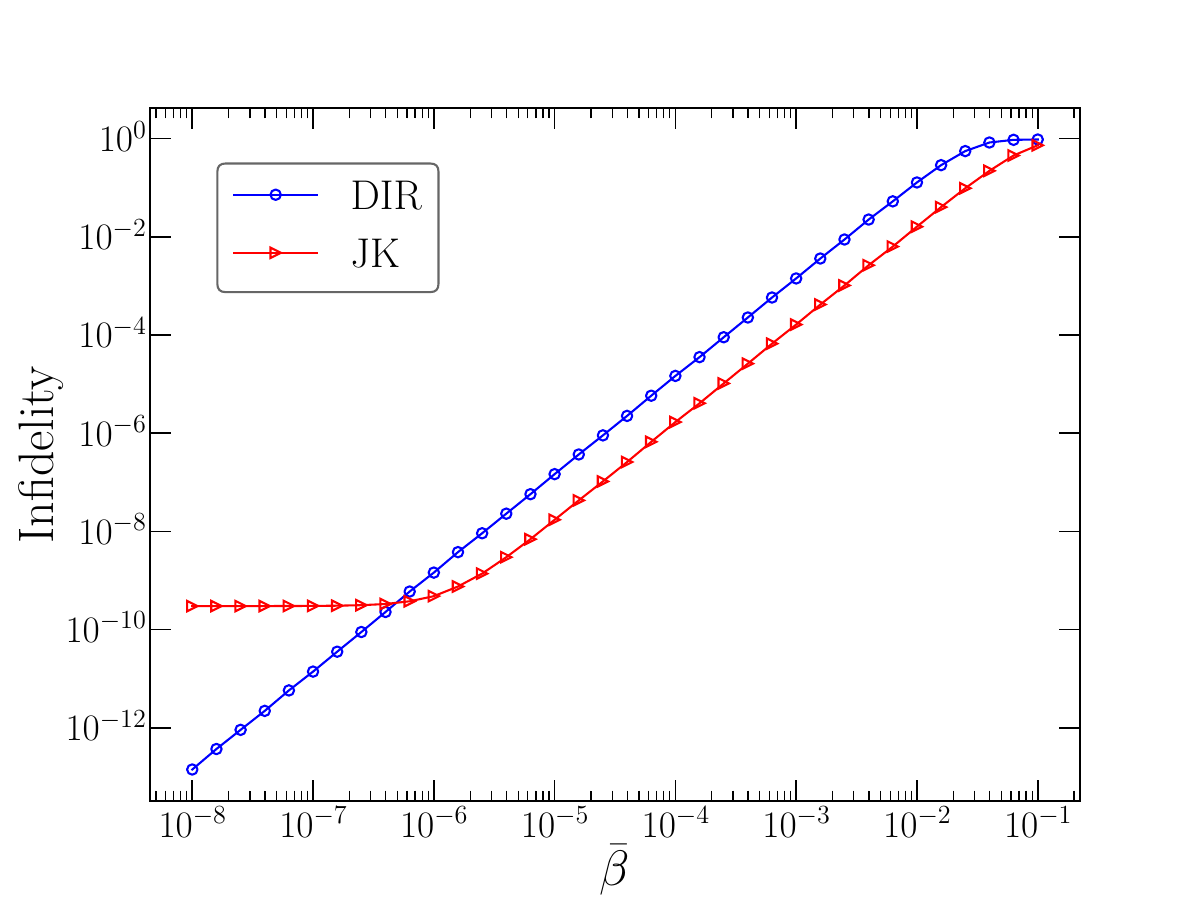}
		\caption{Comparison of the infidelity from DIR and JK sequences as functions of (a) the mean value of charge noise $\alpha$, and (b) the mean value of gate crosstalk  $\beta$.}
		\label{fig:noise performance}
	\end{figure}

	The noise robustness of the gate sequences is demonstrated in Fig.~\ref{fig:noise performance}, where panel (a) shows the effect of charge noise and panel (b) illustrates the effect of crosstalk noise. In both panels, the blue curves correspond to the DIR sequence, while the red curves represent the JK sequence. For each mean value of the noise strength parameter ($\alpha$ in (a), $\beta$ in (b)), the infidelity is obtained by averaging over 100 Monte Carlo samples, where the noise values are sampled from their respective normal distributions.
	
	As the noise strength increases, both sequences exhibit a monotonic rise in infidelity, reflecting their sensitivity to control imperfections or crosstalk. Notably, when the noise strength exceeds $10^{-6}$, the JK sequence (red curve) consistently yields lower infidelity than the DIR sequence (blue curve), indicating its superior noise resilience across a wide range of noise levels. In the low-noise regime ($\bar{\alpha}$ or $\bar{\beta} < 10^{-6}$), the infidelity of the JK sequence saturates at approximately $10^{-10}$, which corresponds to the intrinsic limit achieved during the optimization of the 92-pulse JK sequence. This plateau as $\bar{\alpha} \to 0$ or $\bar{\beta} \to 0$ indicates that optimization precision has been reached. In contrast, the DIR sequence does not exhibit such a plateau and continues to decrease with diminishing noise.
	
	In conclusion, the optimized JK sequence not only achieves higher fidelity and greater robustness against both charge noise and crosstalk compared to the DIR sequence, but also features a reduced pulse count and shorter operation time. These advantages make it a more efficient and reliable choice for quantum gate implementation in semiconductor quantum dot systems, showing its potential for robust and scalable quantum computing applications.

	\begin{table}[tb]
		\centering
		\renewcommand{\arraystretch}{1} 
		\setlength{\tabcolsep}{9pt} 
		\caption{Optimal compilation of Toffoli gate: a sequence with 92 \blue{exchange unitaries} in 50 time steps.}
		\begin{tabular}{|c|c|c|c|c|c|c|c|c|}
			\hline
			\text{Step} & $p_{1,2}$ & $p_{2,3}$ & $p_{3,4}$ & $p_{4,5}$ & $p_{5,6}$ & $p_{6,7}$ & $p_{7,8}$ & $p_{8,9}$ \\ \hline
			1 & 0 & -0.361356 & 0 & 0 & 0 & 0 & 0 & 0 \\ \hline
			2 & 0 & 0 & 0.46694 & 0 & 0 & 0 & 0 & 0 \\ \hline
			3 & 0 & 0.474996 & 0 & -0.6808 & 0 & 0 & 0 & 0 \\ \hline
			4 & 0 & 0 & 0.463992 & 0 & 0.683111 & 0 & 0 & 0 \\ \hline
			5 & 0 & -0.921766 & 0 & -0.352727 & 0 & 0 & 0 & 0 \\ \hline
			6 & 0 & 0 & -0.647173 & 0 & -1.29074 & 0 & 0 & 0 \\ \hline
			7 & 0 & 0 & 0 & -0.783947 & 0 & 0 & 0 & 0 \\ \hline
			8 & 0 & 0 & 0.289479 & 0 & 0.463229 & 0 & 0 & 0 \\ \hline
			9 & 0 & 0.23616 & 0 & -1.03236 & 0 & 0 & 0 & 0 \\ \hline
			10 & 0 & 0 & -0.517465 & 0 & -0.341 & 0 & 0 & 0 \\ \hline
			11 & 0 & 0.445978 & 0 & -0.591699 & 0 & 0 & 0 & 0 \\ \hline
			12 & 0 & 0 & 0.757532 & 0 & 1.29282 & 0 & 0 & 0 \\ \hline
			13 & 0 & 0 & 0 & -0.580782 & 0 & 0 & 0 & 0 \\ \hline
			14 & 0 & 0 & 0 & 0 & -0.355991 & 0 & 0 & 0 \\ \hline
			15 & 0 & 0 & 0 & -0.547022 & 0 & 0 & 0 & 0 \\ \hline
			16 & 0 & 0 & -0.76173 & 0 & 0 & 0 & -0.401742 & 0 \\ \hline
			17 & 0 & 0.400511 & 0 & 0 & 0 & 0.561134 & 0 & -0.985982 \\ \hline
			18 & 0 & 0 & -0.490934 & 0 & 0 & 0 & -0.49032 & 0 \\ \hline
			19 & 0 & 0 & 0 & 0.613639 & 0 & -0.240236 & 0 & 0.782226 \\ \hline
			20 & 0 & 0 & -0.562063 & 0 & 0 & 0 & -0.574828 & 0 \\ \hline
			21 & 0 & 0 & 0 & 0.708219 & 0 & -1.47277 & 0 & -0.0251349 \\ \hline
			22 & 0 & 0 & 0.448977 & 0 & -0.754719 & 0 & 0 & 0 \\ \hline
			23 & 0 & 0.780178 & 0 & 0 & 0 & -0.409208 & 0 & 0 \\ \hline
			24 & 0 & 0 & -0.279016 & 0 & 0 & 0 & -1.07902 & 0 \\ \hline
			25 & 0 & -0.585912 & 0 & 0 & 0 & -0.313122 & 0 & 0 \\ \hline
			26 & 0 & 0 & -1.07795 & 0 & 0.812474 & 0 & 0.370976 & 0 \\ \hline
			27 & 0 & 0.768569 & 0 & 0 & 0 & 0.20101 & 0 & -1.44753 \\ \hline
			28 & 0 & 0 & 0 & 0 & 0 & 0 & -0.571905 & 0 \\ \hline
			29 & 0 & 0 & 0 & 0 & 0 & 1.31895 & 0 & 0.349844 \\ \hline
			30 & 0 & 0 & 0 & 0 & 0.724231 & 0 & -0.647533 & 0 \\ \hline
			31 & 0 & 0 & 0 & 0 & 0 & 1.8308 & 0 & 0 \\ \hline
			32 & 0 & 0 & 0 & 0 & 0 & 0 & 1.34169 & 0 \\ \hline
			33 & 0 & 0 & 0 & 0 & 0 & 0 & 0 & 0.428598 \\ \hline
			34 & 0 & 0 & 0 & 0 & 0.481629 & 0 & -0.807275 & 0 \\ \hline
			35 & 0 & 0 & 0 & 0 & 0 & -1.59216 & 0 & 0 \\ \hline
			36 & 0 & 0 & 0 & 0 & 0.650856 & 0 & -0.0284115 & 0 \\ \hline
			37 & 0 & 0 & 0 & -0.692437 & 0 & 1.29184 & 0 & 0.45061 \\ \hline
			38 & 0 & 0 & 0.51052 & 0 & 0 & 0 & -0.529812 & 0 \\ \hline
			39 & 0 & 0.610313 & 0 & -0.0895998 & 0 & 0.577097 & 0 & 0 \\ \hline
			40 & 0 & 0 & -0.538592 & 0 & -0.649904 & 0 & 1.46932 & 0 \\ \hline
			41 & 0 & 0.6718 & 0 & -0.648449 & 0 & 0 & 0 & 0 \\ \hline
			42 & 0 & 0 & 0.586782 & 0 & -0.353683 & 0 & 0 & 0 \\ \hline
			43 & 0 & 0.669013 & 0 & 0.378554 & 0 & 0 & 0 & 0 \\ \hline
			44 & 0 & 0 & 0 & 0 & -1.18546 & 0 & 0 & 0 \\ \hline
			45 & 0 & 0 & 0 & -1.44203 & 0 & 0 & 0 & 0 \\ \hline
			46 & 0 & 0 & -1.01647 & 0 & 0 & 0 & 0 & 0 \\ \hline
			47 & 0 & 0 & 0 & -0.617632 & 0 & 0 & 0 & 0 \\ \hline
			48 & 0 & 0 & 0.823515 & 0 & -0.376096 & 0 & 0 & 0 \\ \hline
			49 & 0 & -0.475713 & 0 & 1.31782 & 0 & 0 & 0 & 0 \\ \hline
			50 & 0 & 0 & -0.556544 & 0 & 0 & 0 & 0 & 0 \\ \hline
		\end{tabular}\label{tab:JK sequence}
	\end{table}
		
	\section{Conclusion}\label{sec:concl}

In this work, we systematically constructed the angular momentum structure of a three 3-spin EO qubit system and identified the DFS bases. By reformulating the gate implementation problem as a quantum optimal control task, we utilized Krotov's method and its improved JK algorithm to discover a globally optimized 92-pulse sequence for the Toffoli gate. This sequence is significantly shorter than the conventional direct decomposition using single- and two-qubit gates. The noise performance of the JK-optimized sequence was analyzed under charge noise and crosstalk, demonstrating superior robustness compared to the DIR approach. Additionally, the reduced pulse count and shorter operation time further enhance its practicality, achieving higher computational efficiency and reduced cumulative error.

This study not only provides an efficient and robust solution for the Toffoli gate but also establishes a framework for designing high-fidelity quantum gates in noisy environments. The methodologies developed here can be extended to other multi-qubit gates, contributing to the advancement of scalable and fault-tolerant quantum computing. The results demonstrate the potential of optimal control techniques in addressing hardware-specific challenges and optimizing quantum gate implementations.

More broadly, this work reveals a fundamental challenge in EO qubit design: the very encoding scheme that enables universal quantum computation using only exchange interactions, by mapping logical qubits into three-spin total angular momentum subspaces, also introduces substantial complexity into gate construction, particularly for multi-qubit gates.

For instance, even the most efficient two-qubit CNOT sequence (e.g., the Fong-Wandzura construction) already exhibits significant operational overhead. Our analysis of the Toffoli gate shows that such complexity scales unfavorably with system size, rendering na\"ive decompositions using only single- and two-qubit gates highly inefficient. This highlights an open challenge in EO architectures: how to maintain the robustness of exchange-only encoding while achieving efficient, compact realizations of multi-qubit gates. Our JK algorithm provides a promising path forward, but further advances in control design and circuit optimization will be crucial for the long-term scalability of EO-based quantum computing.

	\section*{Acknowledgment}

This work is supported by the National Natural Science Foundation of China (Grant Nos. 11874312 and 12474489), the Research Grants Council of Hong Kong (CityU 11304920), Shenzhen Fundamental Research Program (Grant No. JCYJ20240813153139050), the Guangdong Provincial Quantum Science Strategic Initiative (Grant Nos. GDZX2203001, GDZX2403001), and the Innovation Program for Quantum Science and Technology (Grant No. 2021ZD0302300).

\section*{DATA AVAILABILITY}
The data that support the findings of this article are openly available \cite{data}.

\vfill
	\appendix
	\section{Compilation of Toffoli gate into exchange unitaries by JK algorithm}\label{appx:sequence}

	In this appendix, we show the detail of our results. The unitary in step $k$ is given by Eq.~\eqref{eq:Ukstep}. The compilation of Toffoli gate into exchange unitaries by JK algorithm, which consists of 92 exchange unitaries and 50 time steps is explicitly given in Table~\ref{tab:JK sequence}. \blue{The pulse sequence is described by a set of optimized parameters $p$ for each segment (cf. Eq.~\eqref{eq:fundamentalblock}). This parameter reflects the magnitude of the exchange operation in each segment, integrating both the interaction strength and duration.}
	
%


\onecolumngrid
\vspace{1cm}

\begin{center}
	{\bf\large Supplemental Material}
\end{center}
\vspace{0.5cm}

\setcounter{secnumdepth}{3}  
\setcounter{equation}{0}
\setcounter{figure}{0}
\setcounter{table}{0}
\setcounter{section}{0}

\renewcommand{\theequation}{S-\arabic{equation}}
\renewcommand{\thefigure}{S\arabic{figure}}
\renewcommand{\thetable}{S-\Roman{table}}
\renewcommand\figurename{Supplementary Figure}
\renewcommand\tablename{Supplementary Table}
\newcommand\citetwo[2]{[S\citealp{#1}, S\citealp{#2}]}
\newcommand\citecite[2]{[\citealp{#1}, S\citealp{#2}]}


\makeatletter \renewcommand\@biblabel[1]{[S#1]} \makeatother

\makeatletter \renewcommand\@biblabel[1]{[S#1]} \makeatother


\onecolumngrid

This supplemental material provides additional technical details to support the results presented in the main text. We first present the 90 basis states for the three-qubit system. Next, we explicitly list the ``uncompressed sequence'', a 55-time-step pulse sequence that implements the Toffoli gate using the Krotov algorithm, serving also as the starting point for the Jenga-Krotov (JK) optimization. These materials aim to enhance transparency and reproducibility of our approach and to support future efforts in efficient multi-qubit gate compilation for exchange-only (EO) qubit systems.

\section{Basis states of three-EO-qubit system}\label{app:states}

In this section, we give the complete set of 90 bases states for the three-EO-qubit system. To facilitate the discussion, we use a single digit in the ket notation to indicate single-qubit states, namely $\left|1\right\rangle$ through $\left|8\right\rangle$ are the same as what is given in Eq.~\eqref{eq:8eigenstates} of the main text. We use two digits to indicate two-qubit states, namely $\left|01\right\rangle$ through $\left|64\right\rangle$. The three-qubit states are leveled using three digits in the ket, i.e. $\left|001\right\rangle$ to $\left|090\right\rangle$. Note that $\left|1\right\rangle$, $\left|01\right\rangle$ and $\left|001\right\rangle$ are all different.

For three EO qubits, the quantum numbers are ordered following the convention,
\begin{equation}
	\left|S_{\mathrm{tot}}^{(9)},S_{z,\mathrm{tot}}^{(9)},S_{A,B},S_A,S_B,S_C,S_{A,1,2},S_{B,1,2},S_{C,1,2}\right\rangle,
\end{equation}
then all 90 basis states are given as follows:
\begin{align*}
	&S_{\mathrm{tot}}^{(9)}=\frac{1}{2},S_{z,\mathrm{tot}}^{(9)}=\frac{1}{2},S_{A,B}=0,S_C=\frac{1}{2}\\
	&\left|001\right\rangle=\left|\frac{1}{2},\frac{1}{2},0,\frac{1}{2},\frac{1}{2},\frac{1}{2},0,0,0\right\rangle=\left|01\right\rangle\left|1\right\rangle\\
	&\left|002\right\rangle=\left|\frac{1}{2},\frac{1}{2},0,\frac{1}{2},\frac{1}{2},\frac{1}{2},0,0,1\right\rangle=\left|01\right\rangle\left|3\right\rangle\\
	&\left|003\right\rangle=\left|\frac{1}{2},\frac{1}{2},0,\frac{1}{2},\frac{1}{2},\frac{1}{2},0,1,0\right\rangle=\left|02\right\rangle\left|1\right\rangle\\
	&\left|004\right\rangle=\left|\frac{1}{2},\frac{1}{2},0,\frac{1}{2},\frac{1}{2},\frac{1}{2},0,1,1\right\rangle=\left|02\right\rangle\left|3\right\rangle\\
	&\left|005\right\rangle=\left|\frac{1}{2},\frac{1}{2},0,\frac{1}{2},\frac{1}{2},\frac{1}{2},1,0,0\right\rangle=\left|03\right\rangle\left|1\right\rangle\\
	&\left|006\right\rangle=\left|\frac{1}{2},\frac{1}{2},0,\frac{1}{2},\frac{1}{2},\frac{1}{2},1,0,1\right\rangle=\left|03\right\rangle\left|3\right\rangle\\
	&\left|007\right\rangle=\left|\frac{1}{2},\frac{1}{2},0,\frac{1}{2},\frac{1}{2},\frac{1}{2},1,1,0\right\rangle=\left|04\right\rangle\left|1\right\rangle\\
	&\left|008\right\rangle=\left|\frac{1}{2},\frac{1}{2},0,\frac{1}{2},\frac{1}{2},\frac{1}{2},1,1,1\right\rangle=\left|04\right\rangle\left|3\right\rangle\\
	&\left|009\right\rangle=\left|\frac{1}{2},\frac{1}{2},0,\frac{3}{2},\frac{3}{2},\frac{1}{2},1,1,0\right\rangle=\left|05\right\rangle\left|1\right\rangle\\
	&\left|010\right\rangle=\left|\frac{1}{2},\frac{1}{2},0,\frac{3}{2},\frac{3}{2},\frac{1}{2},1,1,1\right\rangle=\left|05\right\rangle\left|3\right\rangle\\
	&S_{\mathrm{tot}}^{(9)}=\frac{1}{2},S_{z,\mathrm{tot}}^{(9)}=\frac{1}{2},S_{A,B}=1,S_C=\frac{1}{2}\\
	&\left|011\right\rangle=\left|\frac{1}{2},\frac{1}{2},1,\frac{1}{2},\frac{1}{2},\frac{1}{2},0,0,0\right\rangle=\sqrt{\frac{2}{3}}\left|24\right\rangle\left|2\right\rangle-\sqrt{\frac{1}{3}}\left|15\right\rangle\left|1\right\rangle\\
	&\left|012\right\rangle=\left|\frac{1}{2},\frac{1}{2},1,\frac{1}{2},\frac{1}{2},\frac{1}{2},0,0,1\right\rangle=\sqrt{\frac{2}{3}}\left|24\right\rangle\left|4\right\rangle-\sqrt{\frac{1}{3}}\left|15\right\rangle\left|3\right\rangle\\
	&\left|013\right\rangle=\left|\frac{1}{2},\frac{1}{2},1,\frac{1}{2},\frac{1}{2},\frac{1}{2},0,1,0\right\rangle=\sqrt{\frac{2}{3}}\left|25\right\rangle\left|2\right\rangle-\sqrt{\frac{1}{3}}\left|16\right\rangle\left|1\right\rangle\\
	&\left|014\right\rangle=\left|\frac{1}{2},\frac{1}{2},1,\frac{1}{2},\frac{1}{2},\frac{1}{2},0,1,1\right\rangle=\sqrt{\frac{2}{3}}\left|25\right\rangle\left|4\right\rangle-\sqrt{\frac{1}{3}}\left|16\right\rangle\left|3\right\rangle\\
	&\left|015\right\rangle=\left|\frac{1}{2},\frac{1}{2},1,\frac{1}{2},\frac{1}{2},\frac{1}{2},1,0,0\right\rangle=\sqrt{\frac{2}{3}}\left|26\right\rangle\left|2\right\rangle-\sqrt{\frac{1}{3}}\left|17\right\rangle\left|1\right\rangle\\
	&\left|016\right\rangle=\left|\frac{1}{2},\frac{1}{2},1,\frac{1}{2},\frac{1}{2},\frac{1}{2},1,0,1\right\rangle=\sqrt{\frac{2}{3}}\left|26\right\rangle\left|4\right\rangle-\sqrt{\frac{1}{3}}\left|17\right\rangle\left|3\right\rangle\\
	&\left|017\right\rangle=\left|\frac{1}{2},\frac{1}{2},1,\frac{1}{2},\frac{1}{2},\frac{1}{2},1,1,0\right\rangle=\sqrt{\frac{2}{3}}\left|27\right\rangle\left|2\right\rangle-\sqrt{\frac{1}{3}}\left|18\right\rangle\left|1\right\rangle\\
	&\left|018\right\rangle=\left|\frac{1}{2},\frac{1}{2},1,\frac{1}{2},\frac{1}{2},\frac{1}{2},1,1,1\right\rangle=\sqrt{\frac{2}{3}}\left|27\right\rangle\left|4\right\rangle-\sqrt{\frac{1}{3}}\left|18\right\rangle\left|3\right\rangle\\
	&\left|019\right\rangle=\left|\frac{1}{2},\frac{1}{2},1,\frac{1}{2},\frac{3}{2},\frac{1}{2},0,1,0\right\rangle=\sqrt{\frac{2}{3}}\left|28\right\rangle\left|2\right\rangle-\sqrt{\frac{1}{3}}\left|19\right\rangle\left|1\right\rangle\\
	&\left|020\right\rangle=\left|\frac{1}{2},\frac{1}{2},1,\frac{1}{2},\frac{3}{2},\frac{1}{2},0,1,1\right\rangle=\sqrt{\frac{2}{3}}\left|28\right\rangle\left|4\right\rangle-\sqrt{\frac{1}{3}}\left|19\right\rangle\left|3\right\rangle\\
	&\left|021\right\rangle=\left|\frac{1}{2},\frac{1}{2},1,\frac{1}{2},\frac{3}{2},\frac{1}{2},1,1,0\right\rangle=\sqrt{\frac{2}{3}}\left|29\right\rangle\left|2\right\rangle-\sqrt{\frac{1}{3}}\left|20\right\rangle\left|1\right\rangle\\
	&\left|022\right\rangle=\left|\frac{1}{2},\frac{1}{2},1,\frac{1}{2},\frac{3}{2},\frac{1}{2},1,1,1\right\rangle=\sqrt{\frac{2}{3}}\left|29\right\rangle\left|4\right\rangle-\sqrt{\frac{1}{3}}\left|20\right\rangle\left|3\right\rangle\\
	&\left|023\right\rangle=\left|\frac{1}{2},\frac{1}{2},1,\frac{3}{2},\frac{1}{2},\frac{1}{2},1,0,0\right\rangle=\sqrt{\frac{2}{3}}\left|30\right\rangle\left|2\right\rangle-\sqrt{\frac{1}{3}}\left|21\right\rangle\left|1\right\rangle\\
	&\left|024\right\rangle=\left|\frac{1}{2},\frac{1}{2},1,\frac{3}{2},\frac{1}{2},\frac{1}{2},1,0,1\right\rangle=\sqrt{\frac{2}{3}}\left|30\right\rangle\left|4\right\rangle-\sqrt{\frac{1}{3}}\left|21\right\rangle\left|3\right\rangle\\
	&\left|025\right\rangle=\left|\frac{1}{2},\frac{1}{2},1,\frac{3}{2},\frac{1}{2},\frac{1}{2},1,1,0\right\rangle=\sqrt{\frac{2}{3}}\left|31\right\rangle\left|2\right\rangle-\sqrt{\frac{1}{3}}\left|22\right\rangle\left|1\right\rangle\\
	&\left|026\right\rangle=\left|\frac{1}{2},\frac{1}{2},1,\frac{3}{2},\frac{1}{2},\frac{1}{2},1,1,1\right\rangle=\sqrt{\frac{2}{3}}\left|31\right\rangle\left|4\right\rangle-\sqrt{\frac{1}{3}}\left|22\right\rangle\left|3\right\rangle\\
	&\left|027\right\rangle=\left|\frac{1}{2},\frac{1}{2},1,\frac{3}{2},\frac{3}{2},\frac{1}{2},1,1,0\right\rangle=\sqrt{\frac{2}{3}}\left|32\right\rangle\left|2\right\rangle-\sqrt{\frac{1}{3}}\left|23\right\rangle\left|1\right\rangle\\
	&\left|028\right\rangle=\left|\frac{1}{2},\frac{1}{2},1,\frac{3}{2},\frac{3}{2},\frac{1}{2},1,1,1\right\rangle=\sqrt{\frac{2}{3}}\left|32\right\rangle\left|4\right\rangle-\sqrt{\frac{1}{3}}\left|23\right\rangle\left|3\right\rangle\\
	&S_{\mathrm{tot}}^{(9)}=\frac{1}{2},S_{z,\mathrm{tot}}^{(9)}=\frac{1}{2},S_{A,B}=1,S_C=\frac{3}{2}\\
	&\left|029\right\rangle=
	\left|\frac{1}{2},\frac{1}{2},1,\frac{1}{2},\frac{1}{2},\frac{3}{2},0,0,1\right\rangle=
	\sqrt{\frac{1}{2}}\left|06\right\rangle\left|5\right\rangle
	-\sqrt{\frac{1}{3}}\left|15\right\rangle\left|6\right\rangle
	+\sqrt{\frac{1}{6}}\left|24\right\rangle\left|7\right\rangle\\
	&\left|030\right\rangle=
	\left|\frac{1}{2},\frac{1}{2},1,\frac{1}{2},\frac{1}{2},\frac{3}{2},0,1,1\right\rangle=
	\sqrt{\frac{1}{2}}\left|07\right\rangle\left|5\right\rangle
	-\sqrt{\frac{1}{3}}\left|16\right\rangle\left|6\right\rangle
	+\sqrt{\frac{1}{6}}\left|25\right\rangle\left|7\right\rangle\\
	&\left|031\right\rangle=
	\left|\frac{1}{2},\frac{1}{2},1,\frac{1}{2},\frac{1}{2},\frac{3}{2},1,0,1\right\rangle=
	\sqrt{\frac{1}{2}}\left|08\right\rangle\left|5\right\rangle
	-\sqrt{\frac{1}{3}}\left|17\right\rangle\left|6\right\rangle
	+\sqrt{\frac{1}{6}}\left|26\right\rangle\left|7\right\rangle\\
	&\left|032\right\rangle=
	\left|\frac{1}{2},\frac{1}{2},1,\frac{1}{2},\frac{1}{2},\frac{3}{2},1,1,1\right\rangle=
	\sqrt{\frac{1}{2}}\left|09\right\rangle\left|5\right\rangle
	-\sqrt{\frac{1}{3}}\left|18\right\rangle\left|6\right\rangle
	+\sqrt{\frac{1}{6}}\left|27\right\rangle\left|7\right\rangle\\
	&\left|033\right\rangle=
	\left|\frac{1}{2},\frac{1}{2},1,\frac{1}{2},\frac{3}{2},\frac{3}{2},0,1,1\right\rangle=
	\sqrt{\frac{1}{2}}\left|10\right\rangle\left|5\right\rangle
	-\sqrt{\frac{1}{3}}\left|19\right\rangle\left|6\right\rangle
	+\sqrt{\frac{1}{6}}\left|28\right\rangle\left|7\right\rangle\\
	&\left|034\right\rangle=
	\left|\frac{1}{2},\frac{1}{2},1,\frac{1}{2},\frac{3}{2},\frac{3}{2},1,1,1\right\rangle=
	\sqrt{\frac{1}{2}}\left|11\right\rangle\left|5\right\rangle
	-\sqrt{\frac{1}{3}}\left|20\right\rangle\left|6\right\rangle
	+\sqrt{\frac{1}{6}}\left|29\right\rangle\left|7\right\rangle\\
	&\left|035\right\rangle=
	\left|\frac{1}{2},\frac{1}{2},1,\frac{3}{2},\frac{1}{2},\frac{3}{2},1,0,1\right\rangle=
	\sqrt{\frac{1}{2}}\left|12\right\rangle\left|5\right\rangle
	-\sqrt{\frac{1}{3}}\left|21\right\rangle\left|6\right\rangle
	+\sqrt{\frac{1}{6}}\left|30\right\rangle\left|7\right\rangle\\
	&\left|036\right\rangle=
	\left|\frac{1}{2},\frac{1}{2},1,\frac{3}{2},\frac{1}{2},\frac{3}{2},1,1,1\right\rangle=
	\sqrt{\frac{1}{2}}\left|13\right\rangle\left|5\right\rangle
	-\sqrt{\frac{1}{3}}\left|22\right\rangle\left|6\right\rangle
	+\sqrt{\frac{1}{6}}\left|31\right\rangle\left|7\right\rangle\\
	&\left|037\right\rangle=
	\left|\frac{1}{2},\frac{1}{2},1,\frac{3}{2},\frac{3}{2},\frac{3}{2},1,1,1\right\rangle=
	\sqrt{\frac{1}{2}}\left|14\right\rangle\left|5\right\rangle
	-\sqrt{\frac{1}{3}}\left|23\right\rangle\left|6\right\rangle
	+\sqrt{\frac{1}{6}}\left|32\right\rangle\left|7\right\rangle\\
	&S_{\mathrm{tot}}^{(9)}=\frac{1}{2},S_{z,\mathrm{tot}}^{(9)}=\frac{1}{2},S_{A,B}=2,S_C=\frac{3}{2}\\
	&\left|038\right\rangle=
	\left|\frac{1}{2},\frac{1}{2},2,\frac{1}{2},\frac{3}{2},\frac{3}{2},0,1,1\right\rangle=
	-\sqrt{\frac{1}{10}}\left|38\right\rangle\left|5\right\rangle
	+\sqrt{\frac{1}{5}}\left|43\right\rangle\left|6\right\rangle
	-\sqrt{\frac{3}{10}}\left|48\right\rangle\left|7\right\rangle
	+\sqrt{\frac{2}{5}}\left|53\right\rangle\left|8\right\rangle\\
	&\left|039\right\rangle=
	\left|\frac{1}{2},\frac{1}{2},2,\frac{1}{2},\frac{3}{2},\frac{3}{2},1,1,1\right\rangle=
	-\sqrt{\frac{1}{10}}\left|39\right\rangle\left|5\right\rangle
	+\sqrt{\frac{1}{5}}\left|44\right\rangle\left|6\right\rangle
	-\sqrt{\frac{3}{10}}\left|49\right\rangle\left|7\right\rangle
	+\sqrt{\frac{2}{5}}\left|54\right\rangle\left|8\right\rangle\\
	&\left|040\right\rangle=
	\left|\frac{1}{2},\frac{1}{2},2,\frac{3}{2},\frac{1}{2},\frac{3}{2},1,0,1\right\rangle=
	-\sqrt{\frac{1}{10}}\left|40\right\rangle\left|5\right\rangle
	+\sqrt{\frac{1}{5}}\left|45\right\rangle\left|6\right\rangle
	-\sqrt{\frac{3}{10}}\left|50\right\rangle\left|7\right\rangle
	+\sqrt{\frac{2}{5}}\left|55\right\rangle\left|8\right\rangle\\
	&\left|041\right\rangle=
	\left|\frac{1}{2},\frac{1}{2},2,\frac{3}{2},\frac{1}{2},\frac{3}{2},1,1,1\right\rangle=
	-\sqrt{\frac{1}{10}}\left|41\right\rangle\left|5\right\rangle
	+\sqrt{\frac{1}{5}}\left|46\right\rangle\left|6\right\rangle
	-\sqrt{\frac{3}{10}}\left|51\right\rangle\left|7\right\rangle
	+\sqrt{\frac{2}{5}}\left|56\right\rangle\left|8\right\rangle\\
	&\left|042\right\rangle=
	\left|\frac{1}{2},\frac{1}{2},2,\frac{3}{2},\frac{3}{2},\frac{3}{2},1,1,1\right\rangle=
	-\sqrt{\frac{1}{10}}\left|42\right\rangle\left|5\right\rangle
	+\sqrt{\frac{1}{5}}\left|47\right\rangle\left|6\right\rangle
	-\sqrt{\frac{3}{10}}\left|52\right\rangle\left|7\right\rangle
	+\sqrt{\frac{2}{5}}\left|57\right\rangle\left|8\right\rangle\\
	&S_{\mathrm{tot}}^{(9)}=\frac{3}{2},S_{z,\mathrm{tot}}^{(9)}=\frac{3}{2},S_{A,B}=1,S_C=\frac{1}{2}\\
	&\left|043\right\rangle=\left|\frac{3}{2},\frac{3}{2},1,\frac{1}{2},\frac{1}{2},\frac{1}{2},0,0,0\right\rangle=\left|24\right\rangle\left|1\right\rangle\\
	&\left|044\right\rangle=\left|\frac{3}{2},\frac{3}{2},1,\frac{1}{2},\frac{1}{2},\frac{1}{2},0,0,1\right\rangle=\left|24\right\rangle\left|3\right\rangle\\
	&\left|045\right\rangle=\left|\frac{3}{2},\frac{3}{2},1,\frac{1}{2},\frac{1}{2},\frac{1}{2},0,1,0\right\rangle=\left|25\right\rangle\left|1\right\rangle\\
	&\left|046\right\rangle=\left|\frac{3}{2},\frac{3}{2},1,\frac{1}{2},\frac{1}{2},\frac{1}{2},0,1,1\right\rangle=\left|25\right\rangle\left|3\right\rangle\\
	&\left|047\right\rangle=\left|\frac{3}{2},\frac{3}{2},1,\frac{1}{2},\frac{1}{2},\frac{1}{2},1,0,0\right\rangle=\left|26\right\rangle\left|1\right\rangle\\
	&\left|048\right\rangle=\left|\frac{3}{2},\frac{3}{2},1,\frac{1}{2},\frac{1}{2},\frac{1}{2},1,0,1\right\rangle=\left|26\right\rangle\left|3\right\rangle\\
	&\left|049\right\rangle=\left|\frac{3}{2},\frac{3}{2},1,\frac{1}{2},\frac{1}{2},\frac{1}{2},1,1,0\right\rangle=\left|27\right\rangle\left|1\right\rangle\\
	&\left|050\right\rangle=\left|\frac{3}{2},\frac{3}{2},1,\frac{1}{2},\frac{1}{2},\frac{1}{2},1,1,1\right\rangle=\left|27\right\rangle\left|3\right\rangle\\
	&\left|051\right\rangle=\left|\frac{3}{2},\frac{3}{2},1,\frac{1}{2},\frac{3}{2},\frac{1}{2},0,1,0\right\rangle=\left|28\right\rangle\left|1\right\rangle\\
	&\left|052\right\rangle=\left|\frac{3}{2},\frac{3}{2},1,\frac{1}{2},\frac{3}{2},\frac{1}{2},0,1,1\right\rangle=\left|28\right\rangle\left|3\right\rangle\\
	&\left|053\right\rangle=\left|\frac{3}{2},\frac{3}{2},1,\frac{1}{2},\frac{3}{2},\frac{1}{2},1,1,0\right\rangle=\left|29\right\rangle\left|1\right\rangle\\
	&\left|054\right\rangle=\left|\frac{3}{2},\frac{3}{2},1,\frac{1}{2},\frac{3}{2},\frac{1}{2},1,1,1\right\rangle=\left|29\right\rangle\left|3\right\rangle\\
	&\left|055\right\rangle=\left|\frac{3}{2},\frac{3}{2},1,\frac{3}{2},\frac{1}{2},\frac{1}{2},1,0,0\right\rangle=\left|30\right\rangle\left|1\right\rangle\\
	&\left|056\right\rangle=\left|\frac{3}{2},\frac{3}{2},1,\frac{3}{2},\frac{1}{2},\frac{1}{2},1,0,1\right\rangle=\left|30\right\rangle\left|3\right\rangle\\
	&\left|057\right\rangle=\left|\frac{3}{2},\frac{3}{2},1,\frac{3}{2},\frac{1}{2},\frac{1}{2},1,1,0\right\rangle=\left|31\right\rangle\left|1\right\rangle\\
	&\left|058\right\rangle=\left|\frac{3}{2},\frac{3}{2},1,\frac{3}{2},\frac{1}{2},\frac{1}{2},1,1,1\right\rangle=\left|31\right\rangle\left|3\right\rangle\\
	&\left|059\right\rangle=\left|\frac{3}{2},\frac{3}{2},1,\frac{3}{2},\frac{3}{2},\frac{1}{2},1,1,0\right\rangle=\left|32\right\rangle\left|1\right\rangle\\
	&\left|060\right\rangle=\left|\frac{3}{2},\frac{3}{2},1,\frac{3}{2},\frac{3}{2},\frac{1}{2},1,1,1\right\rangle=\left|32\right\rangle\left|3\right\rangle\\
	&S_{\mathrm{tot}}^{(9)}=\frac{3}{2},S_{z,\mathrm{tot}}^{(9)}=\frac{3}{2},S_{A,B}=2,S_C=\frac{1}{2}\\
	&\left|061\right\rangle=
	\left|\frac{3}{2},\frac{3}{2},2,\frac{1}{2},\frac{3}{2},\frac{1}{2},0,1,0\right\rangle=
	\sqrt{\frac{4}{5}}\left|53\right\rangle\left|2\right\rangle
	-\sqrt{\frac{1}{5}}\left|48\right\rangle\left|1\right\rangle\\
	&\left|062\right\rangle=
	\left|\frac{3}{2},\frac{3}{2},2,\frac{1}{2},\frac{3}{2},\frac{1}{2},0,1,1\right\rangle=
	\sqrt{\frac{4}{5}}\left|53\right\rangle\left|4\right\rangle
	-\sqrt{\frac{1}{5}}\left|48\right\rangle\left|3\right\rangle\\
	&\left|063\right\rangle=
	\left|\frac{3}{2},\frac{3}{2},2,\frac{1}{2},\frac{3}{2},\frac{1}{2},1,1,0\right\rangle=
	\sqrt{\frac{4}{5}}\left|54\right\rangle\left|2\right\rangle
	-\sqrt{\frac{1}{5}}\left|49\right\rangle\left|1\right\rangle\\
	&\left|064\right\rangle=
	\left|\frac{3}{2},\frac{3}{2},2,\frac{1}{2},\frac{3}{2},\frac{1}{2},1,1,1\right\rangle=
	\sqrt{\frac{4}{5}}\left|54\right\rangle\left|4\right\rangle
	-\sqrt{\frac{1}{5}}\left|49\right\rangle\left|3\right\rangle\\
	&\left|065\right\rangle=
	\left|\frac{3}{2},\frac{3}{2},2,\frac{3}{2},\frac{1}{2},\frac{1}{2},1,0,0\right\rangle=
	\sqrt{\frac{4}{5}}\left|55\right\rangle\left|2\right\rangle
	-\sqrt{\frac{1}{5}}\left|50\right\rangle\left|1\right\rangle\\
	&\left|066\right\rangle=
	\left|\frac{3}{2},\frac{3}{2},2,\frac{3}{2},\frac{1}{2},\frac{1}{2},1,0,1\right\rangle=
	\sqrt{\frac{4}{5}}\left|55\right\rangle\left|4\right\rangle
	-\sqrt{\frac{1}{5}}\left|50\right\rangle\left|3\right\rangle\\
	&\left|067\right\rangle=
	\left|\frac{3}{2},\frac{3}{2},2,\frac{3}{2},\frac{1}{2},\frac{1}{2},1,1,0\right\rangle=
	\sqrt{\frac{4}{5}}\left|56\right\rangle\left|2\right\rangle
	-\sqrt{\frac{1}{5}}\left|51\right\rangle\left|1\right\rangle\\
	&\left|068\right\rangle=
	\left|\frac{3}{2},\frac{3}{2},2,\frac{3}{2},\frac{1}{2},\frac{1}{2},1,1,1\right\rangle=
	\sqrt{\frac{4}{5}}\left|56\right\rangle\left|4\right\rangle
	-\sqrt{\frac{1}{5}}\left|51\right\rangle\left|3\right\rangle\\
	&\left|069\right\rangle=
	\left|\frac{3}{2},\frac{3}{2},2,\frac{3}{2},\frac{3}{2},\frac{1}{2},1,1,0\right\rangle=
	\sqrt{\frac{4}{5}}\left|57\right\rangle\left|2\right\rangle
	-\sqrt{\frac{1}{5}}\left|52\right\rangle\left|1\right\rangle\\
	&\left|070\right\rangle=
	\left|\frac{3}{2},\frac{3}{2},2,\frac{3}{2},\frac{3}{2},\frac{1}{2},1,1,1\right\rangle=
	\sqrt{\frac{4}{5}}\left|57\right\rangle\left|4\right\rangle
	-\sqrt{\frac{1}{5}}\left|52\right\rangle\left|3\right\rangle\\
	&S_{\mathrm{tot}}^{(9)}=\frac{3}{2},S_{z,\mathrm{tot}}^{(9)}=\frac{3}{2},S_{A,B}=0,S_C=\frac{3}{2}\\
	&\left|071\right\rangle=
	\left|\frac{3}{2},\frac{3}{2},0,\frac{1}{2},\frac{1}{2},\frac{3}{2},0,0,1\right\rangle=
	\left|01\right\rangle\left|5\right\rangle\\
	&\left|072\right\rangle=
	\left|\frac{3}{2},\frac{3}{2},0,\frac{1}{2},\frac{1}{2},\frac{3}{2},0,1,1\right\rangle=
	\left|02\right\rangle\left|5\right\rangle\\
	&\left|073\right\rangle=
	\left|\frac{3}{2},\frac{3}{2},0,\frac{1}{2},\frac{1}{2},\frac{3}{2},1,0,1\right\rangle=
	\left|03\right\rangle\left|5\right\rangle\\
	&\left|074\right\rangle=
	\left|\frac{3}{2},\frac{3}{2},0,\frac{1}{2},\frac{1}{2},\frac{3}{2},1,1,1\right\rangle=
	\left|04\right\rangle\left|5\right\rangle\\
	&\left|075\right\rangle=
	\left|\frac{3}{2},\frac{3}{2},0,\frac{3}{2},\frac{3}{2},\frac{3}{2},1,1,1\right\rangle=
	\left|05\right\rangle\left|5\right\rangle\\
	&S_{\mathrm{tot}}^{(9)}=\frac{3}{2},S_{z,\mathrm{tot}}^{(9)}=\frac{3}{2},S_{A,B}=1,S_C=\frac{3}{2}\\
	&\left|076\right\rangle=
	\left|\frac{3}{2},\frac{3}{2},1,\frac{1}{2},\frac{1}{2},\frac{3}{2},0,0,1\right\rangle=
	\sqrt{\frac{3}{5}}\left|15\right\rangle\left|5\right\rangle
	-\sqrt{\frac{2}{5}}\left|24\right\rangle\left|6\right\rangle\\
	&\left|077\right\rangle=
	\left|\frac{3}{2},\frac{3}{2},1,\frac{1}{2},\frac{1}{2},\frac{3}{2},0,1,1\right\rangle=
	\sqrt{\frac{3}{5}}\left|16\right\rangle\left|5\right\rangle
	-\sqrt{\frac{2}{5}}\left|25\right\rangle\left|6\right\rangle\\
	&\left|078\right\rangle=
	\left|\frac{3}{2},\frac{3}{2},1,\frac{1}{2},\frac{1}{2},\frac{3}{2},1,0,1\right\rangle=
	\sqrt{\frac{3}{5}}\left|17\right\rangle\left|5\right\rangle
	-\sqrt{\frac{2}{5}}\left|26\right\rangle\left|6\right\rangle\\
	&\left|079\right\rangle=
	\left|\frac{3}{2},\frac{3}{2},1,\frac{1}{2},\frac{1}{2},\frac{3}{2},1,1,1\right\rangle=
	\sqrt{\frac{3}{5}}\left|18\right\rangle\left|5\right\rangle
	-\sqrt{\frac{2}{5}}\left|27\right\rangle\left|6\right\rangle\\
	&\left|080\right\rangle=
	\left|\frac{3}{2},\frac{3}{2},1,\frac{1}{2},\frac{3}{2},\frac{3}{2},0,1,1\right\rangle=
	\sqrt{\frac{3}{5}}\left|19\right\rangle\left|5\right\rangle
	-\sqrt{\frac{2}{5}}\left|28\right\rangle\left|6\right\rangle\\
	&\left|081\right\rangle=
	\left|\frac{3}{2},\frac{3}{2},1,\frac{1}{2},\frac{3}{2},\frac{3}{2},1,1,1\right\rangle=
	\sqrt{\frac{3}{5}}\left|20\right\rangle\left|5\right\rangle
	-\sqrt{\frac{2}{5}}\left|29\right\rangle\left|6\right\rangle\\
	&\left|082\right\rangle=
	\left|\frac{3}{2},\frac{3}{2},1,\frac{3}{2},\frac{1}{2},\frac{3}{2},1,0,1\right\rangle=
	\sqrt{\frac{3}{5}}\left|21\right\rangle\left|5\right\rangle
	-\sqrt{\frac{2}{5}}\left|30\right\rangle\left|6\right\rangle\\
	&\left|083\right\rangle=
	\left|\frac{3}{2},\frac{3}{2},1,\frac{3}{2},\frac{1}{2},\frac{3}{2},1,1,1\right\rangle=
	\sqrt{\frac{3}{5}}\left|22\right\rangle\left|5\right\rangle
	-\sqrt{\frac{2}{5}}\left|31\right\rangle\left|6\right\rangle\\
	&\left|084\right\rangle=
	\left|\frac{3}{2},\frac{3}{2},1,\frac{3}{2},\frac{3}{2},\frac{3}{2},1,1,1\right\rangle=
	\sqrt{\frac{3}{5}}\left|23\right\rangle\left|5\right\rangle
	-\sqrt{\frac{2}{5}}\left|32\right\rangle\left|6\right\rangle\\
	&S_{\mathrm{tot}}^{(9)}=\frac{3}{2},S_{z,\mathrm{tot}}^{(9)}=\frac{3}{2},S_{A,B}=2,S_C=\frac{3}{2}\\
	&\left|085\right\rangle=
	\left|\frac{3}{2},\frac{3}{2},2,\frac{1}{2},\frac{3}{2},\frac{3}{2},0,1,1\right\rangle=
	\sqrt{\frac{1}{5}}\left|43\right\rangle\left|5\right\rangle
	-\sqrt{\frac{2}{5}}\left|48\right\rangle\left|6\right\rangle
	+\sqrt{\frac{2}{5}}\left|53\right\rangle\left|7\right\rangle\\
	&\left|086\right\rangle=
	\left|\frac{3}{2},\frac{3}{2},2,\frac{1}{2},\frac{3}{2},\frac{3}{2},1,1,1\right\rangle=
	\sqrt{\frac{1}{5}}\left|44\right\rangle\left|5\right\rangle
	-\sqrt{\frac{2}{5}}\left|49\right\rangle\left|6\right\rangle
	+\sqrt{\frac{2}{5}}\left|54\right\rangle\left|7\right\rangle\\
	&\left|087\right\rangle=
	\left|\frac{3}{2},\frac{3}{2},2,\frac{3}{2},\frac{1}{2},\frac{3}{2},1,0,1\right\rangle=
	\sqrt{\frac{1}{5}}\left|45\right\rangle\left|5\right\rangle
	-\sqrt{\frac{2}{5}}\left|50\right\rangle\left|6\right\rangle
	+\sqrt{\frac{2}{5}}\left|55\right\rangle\left|7\right\rangle\\
	&\left|088\right\rangle=
	\left|\frac{3}{2},\frac{3}{2},2,\frac{3}{2},\frac{1}{2},\frac{3}{2},1,1,1\right\rangle=
	\sqrt{\frac{1}{5}}\left|46\right\rangle\left|5\right\rangle
	-\sqrt{\frac{2}{5}}\left|51\right\rangle\left|6\right\rangle
	+\sqrt{\frac{2}{5}}\left|56\right\rangle\left|7\right\rangle\\
	&\left|089\right\rangle=
	\left|\frac{3}{2},\frac{3}{2},2,\frac{3}{2},\frac{3}{2},\frac{3}{2},1,1,1\right\rangle=
	\sqrt{\frac{1}{5}}\left|47\right\rangle\left|5\right\rangle
	-\sqrt{\frac{2}{5}}\left|52\right\rangle\left|6\right\rangle
	+\sqrt{\frac{2}{5}}\left|57\right\rangle\left|7\right\rangle\\
	&S_{\mathrm{tot}}^{(9)}=\frac{3}{2},S_{z,\mathrm{tot}}^{(9)}=\frac{3}{2},S_{A,B}=3,S_C=\frac{3}{2}\\
	&\left|090\right\rangle=
	\left|\frac{3}{2},\frac{3}{2},3,\frac{3}{2},\frac{3}{2},\frac{3}{2},1,1,1\right\rangle=
	-\sqrt{\frac{1}{35}}\left|61\right\rangle\left|5\right\rangle
	+\sqrt{\frac{4}{35}}\left|62\right\rangle\left|6\right\rangle
	-\sqrt{\frac{10}{35}}\left|63\right\rangle\left|7\right\rangle
	+\sqrt{\frac{20}{35}}\left|64\right\rangle\left|8\right\rangle.
\end{align*}
$\left|01\right\rangle$ to $\left|64\right\rangle$ are two-qubit basis states, which are given in terms of single-qubit bases under quantum numbers ordering as $\left|S_{\mathrm{tot}}^{(6)},S_{z,\mathrm{tot}}^{(6)},S_A,S_B,S_{A,1,2},S_{B,1,2}\right\rangle$:
\begin{align*}
	&S_{\mathrm{tot}}^{(6)}=0,S_{z,\mathrm{tot}}^{(6)}=0\\
	&\left|01\right\rangle=\left|0,0,\frac{1}{2},\frac{1}{2},0,0\right\rangle=
	\sqrt{\frac{1}{2}}\left|1\right\rangle\left|2\right\rangle-\sqrt{\frac{1}{2}}\left|2\right\rangle\left|1\right\rangle\\
	&\left|02\right\rangle=\left|0,0,\frac{1}{2},\frac{1}{2},0,1\right\rangle=
	\sqrt{\frac{1}{2}}\left|1\right\rangle\left|4\right\rangle-\sqrt{\frac{1}{2}}\left|2\right\rangle\left|3\right\rangle\\
	&\left|03\right\rangle=\left|0,0,\frac{1}{2},\frac{1}{2},1,0\right\rangle=
	\sqrt{\frac{1}{2}}\left|3\right\rangle\left|2\right\rangle-\sqrt{\frac{1}{2}}\left|4\right\rangle\left|1\right\rangle\\
	&\left|04\right\rangle=\left|0,0,\frac{1}{2},\frac{1}{2},1,1\right\rangle=
	\sqrt{\frac{1}{2}}\left|3\right\rangle\left|4\right\rangle-\sqrt{\frac{1}{2}}\left|4\right\rangle\left|3\right\rangle\\
	&\left|05\right\rangle=\left|0,0,\frac{3}{2},\frac{3}{2},1,1\right\rangle=
	\sqrt{\frac{1}{4}}\left|5\right\rangle\left|8\right\rangle-\sqrt{\frac{1}{4}}\left|6\right\rangle\left|7\right\rangle+
	\sqrt{\frac{1}{4}}\left|7\right\rangle\left|6\right\rangle-\sqrt{\frac{1}{4}}\left|8\right\rangle\left|5\right\rangle\\
	&S_{\mathrm{tot}}^{(6)}=1,S_{z,\mathrm{tot}}^{(6)}=-1\\
	&\left|06\right\rangle=\left|1,-1,\frac{1}{2},\frac{1}{2},0,0\right\rangle=\left|2\right\rangle\left|2\right\rangle\\
	&\left|07\right\rangle=\left|1,-1,\frac{1}{2},\frac{1}{2},0,1\right\rangle=\left|2\right\rangle\left|4\right\rangle\\
	&\left|08\right\rangle=\left|1,-1,\frac{1}{2},\frac{1}{2},1,0\right\rangle=\left|4\right\rangle\left|2\right\rangle\\
	&\left|09\right\rangle=\left|1,-1,\frac{1}{2},\frac{1}{2},1,1\right\rangle=\left|4\right\rangle\left|4\right\rangle\\
	&\left|10\right\rangle=\left|1,-1,\frac{1}{2},\frac{3}{2},0,1\right\rangle=
	\sqrt{\frac{1}{4}}\left|2\right\rangle\left|7\right\rangle-\sqrt{\frac{3}{4}}\left|1\right\rangle\left|8\right\rangle\\
	&\left|11\right\rangle=\left|1,-1,\frac{1}{2},\frac{3}{2},1,1\right\rangle=
	\sqrt{\frac{1}{4}}\left|4\right\rangle\left|7\right\rangle-\sqrt{\frac{3}{4}}\left|3\right\rangle\left|8\right\rangle\\
	&\left|12\right\rangle=\left|1,-1,\frac{3}{2},\frac{1}{2},1,0\right\rangle=
	\sqrt{\frac{1}{4}}\left|7\right\rangle\left|2\right\rangle-\sqrt{\frac{3}{4}}\left|8\right\rangle\left|1\right\rangle\\
	&\left|13\right\rangle=\left|1,-1,\frac{3}{2},\frac{1}{2},1,1\right\rangle=
	\sqrt{\frac{1}{4}}\left|7\right\rangle\left|4\right\rangle-\sqrt{\frac{3}{4}}\left|8\right\rangle\left|3\right\rangle\\
	&\left|14\right\rangle=\left|1,-1,\frac{3}{2},\frac{3}{2},1,1\right\rangle=\sqrt{\frac{3}{10}}\left|6\right\rangle\left|8\right\rangle-
	\sqrt{\frac{2}{5}}\left|7\right\rangle\left|7\right\rangle+\sqrt{\frac{3}{10}}\left|8\right\rangle\left|6\right\rangle\\
	&S_{\mathrm{tot}}^{(6)}=1,S_{z,\mathrm{tot}}^{(6)}=0\\
	&\left|15\right\rangle=\left|1,0,\frac{1}{2},\frac{1}{2},0,0\right\rangle=
	\sqrt{\frac{1}{2}}\left|1\right\rangle\left|2\right\rangle+\sqrt{\frac{1}{2}}\left|2\right\rangle\left|1\right\rangle\\
	&\left|16\right\rangle=\left|1,0,\frac{1}{2},\frac{1}{2},0,1\right\rangle=
	\sqrt{\frac{1}{2}}\left|1\right\rangle\left|4\right\rangle+\sqrt{\frac{1}{2}}\left|2\right\rangle\left|3\right\rangle\\
	&\left|17\right\rangle=\left|1,0,\frac{1}{2},\frac{1}{2},1,0\right\rangle=
	\sqrt{\frac{1}{2}}\left|3\right\rangle\left|2\right\rangle+\sqrt{\frac{1}{2}}\left|4\right\rangle\left|1\right\rangle\\
	&\left|18\right\rangle=\left|1,0,\frac{1}{2},\frac{1}{2},1,1\right\rangle=
	\sqrt{\frac{1}{2}}\left|3\right\rangle\left|4\right\rangle+\sqrt{\frac{1}{2}}\left|4\right\rangle\left|3\right\rangle\\
	&\left|19\right\rangle=\left|1,0,\frac{1}{2},\frac{3}{2},0,1\right\rangle=
	\sqrt{\frac{1}{2}}\left|2\right\rangle\left|6\right\rangle-\sqrt{\frac{1}{2}}\left|1\right\rangle\left|7\right\rangle\\
	&\left|20\right\rangle=\left|1,0,\frac{1}{2},\frac{3}{2},1,1\right\rangle=
	\sqrt{\frac{1}{2}}\left|4\right\rangle\left|6\right\rangle-\sqrt{\frac{1}{2}}\left|3\right\rangle\left|7\right\rangle\\
	&\left|21\right\rangle=\left|1,0,\frac{3}{2},\frac{1}{2},1,0\right\rangle=
	\sqrt{\frac{1}{2}}\left|6\right\rangle\left|2\right\rangle-\sqrt{\frac{1}{2}}\left|7\right\rangle\left|1\right\rangle\\
	&\left|22\right\rangle=\left|1,0,\frac{3}{2},\frac{1}{2},1,1\right\rangle=
	\sqrt{\frac{1}{2}}\left|6\right\rangle\left|4\right\rangle-\sqrt{\frac{1}{2}}\left|7\right\rangle\left|3\right\rangle\\
	&\left|23\right\rangle=\left|1,0,\frac{3}{2},\frac{3}{2},1,1\right\rangle=
	\sqrt{\frac{9}{20}}\left|5\right\rangle\left|8\right\rangle-\sqrt{\frac{1}{20}}\left|6\right\rangle\left|7\right\rangle-
	\sqrt{\frac{1}{20}}\left|7\right\rangle\left|6\right\rangle+\sqrt{\frac{9}{20}}\left|8\right\rangle\left|5\right\rangle\\
	&S_{\mathrm{tot}}^{(6)}=1,S_{z,\mathrm{tot}}^{(6)}=1\\
	&\left|24\right\rangle=\left|1,1,\frac{1}{2},\frac{1}{2},0,0\right\rangle=\left|1\right\rangle\left|1\right\rangle\\
	&\left|25\right\rangle=\left|1,1,\frac{1}{2},\frac{1}{2},0,1\right\rangle=\left|1\right\rangle\left|3\right\rangle\\
	&\left|26\right\rangle=\left|1,1,\frac{1}{2},\frac{1}{2},1,0\right\rangle=\left|3\right\rangle\left|1\right\rangle\\
	&\left|27\right\rangle=\left|1,1,\frac{1}{2},\frac{1}{2},1,1\right\rangle=\left|3\right\rangle\left|3\right\rangle\\
	&\left|28\right\rangle=\left|1,1,\frac{1}{2},\frac{3}{2},0,1\right\rangle=
	\sqrt{\frac{3}{4}}\left|2\right\rangle\left|5\right\rangle-\sqrt{\frac{1}{4}}\left|1\right\rangle\left|6\right\rangle\\
	&\left|29\right\rangle=\left|1,1,\frac{1}{2},\frac{3}{2},1,1\right\rangle=
	\sqrt{\frac{3}{4}}\left|4\right\rangle\left|5\right\rangle-\sqrt{\frac{1}{4}}\left|3\right\rangle\left|6\right\rangle\\
	&\left|30\right\rangle=\left|1,1,\frac{3}{2},\frac{1}{2},1,0\right\rangle=
	\sqrt{\frac{3}{4}}\left|5\right\rangle\left|2\right\rangle-\sqrt{\frac{1}{4}}\left|6\right\rangle\left|1\right\rangle\\
	&\left|31\right\rangle=\left|1,1,\frac{3}{2},\frac{1}{2},1,1\right\rangle=
	\sqrt{\frac{3}{4}}\left|5\right\rangle\left|4\right\rangle-\sqrt{\frac{1}{4}}\left|6\right\rangle\left|3\right\rangle\\
	&\left|32\right\rangle=\left|1,1,\frac{3}{2},\frac{3}{2},1,1\right\rangle=\sqrt{\frac{3}{10}}\left|5\right\rangle\left|7\right\rangle-
	\sqrt{\frac{2}{5}}\left|6\right\rangle\left|6\right\rangle+\sqrt{\frac{3}{10}}\left|7\right\rangle\left|5\right\rangle\\
	&S_{\mathrm{tot}}^{(6)}=2,S_{z,\mathrm{tot}}^{(6)}=-2\\
	&\left|33\right\rangle=\left|2,-2,\frac{1}{2},\frac{3}{2},0,1\right\rangle=\left|2\right\rangle\left|8\right\rangle\\
	&\left|34\right\rangle=\left|2,-2,\frac{1}{2},\frac{3}{2},1,1\right\rangle=\left|4\right\rangle\left|8\right\rangle\\
	&\left|35\right\rangle=\left|2,-2,\frac{3}{2},\frac{1}{2},1,0\right\rangle=\left|8\right\rangle\left|2\right\rangle\\
	&\left|36\right\rangle=\left|2,-2,\frac{3}{2},\frac{1}{2},1,1\right\rangle=\left|8\right\rangle\left|4\right\rangle\\
	&\left|37\right\rangle=\left|2,-2,\frac{3}{2},\frac{3}{2},1,1\right\rangle=
	\sqrt{\frac{1}{2}}\left|7\right\rangle\left|8\right\rangle-\sqrt{\frac{1}{2}}\left|8\right\rangle\left|7\right\rangle\\
	&S_{\mathrm{tot}}^{(6)}=2,S_{z,\mathrm{tot}}^{(6)}=-1\\
	&\left|38\right\rangle=\left|2,-1,\frac{1}{2},\frac{3}{2},0,1\right\rangle=
	\sqrt{\frac{3}{4}}\left|2\right\rangle\left|7\right\rangle+\sqrt{\frac{1}{4}}\left|1\right\rangle\left|8\right\rangle\\
	&\left|39\right\rangle=\left|2,-1,\frac{1}{2},\frac{3}{2},1,1\right\rangle=
	\sqrt{\frac{3}{4}}\left|4\right\rangle\left|7\right\rangle+\sqrt{\frac{1}{4}}\left|3\right\rangle\left|8\right\rangle\\
	&\left|40\right\rangle=\left|2,-1,\frac{3}{2},\frac{1}{2},1,0\right\rangle=
	\sqrt{\frac{3}{4}}\left|7\right\rangle\left|2\right\rangle+\sqrt{\frac{1}{4}}\left|8\right\rangle\left|1\right\rangle\\
	&\left|41\right\rangle=\left|2,-1,\frac{3}{2},\frac{1}{2},1,1\right\rangle=
	\sqrt{\frac{3}{4}}\left|7\right\rangle\left|4\right\rangle+\sqrt{\frac{1}{4}}\left|8\right\rangle\left|3\right\rangle\\
	&\left|42\right\rangle=\left|2,-1,\frac{3}{2},\frac{3}{2},1,1\right\rangle=
	\sqrt{\frac{1}{2}}\left|6\right\rangle\left|8\right\rangle-\sqrt{\frac{1}{2}}\left|8\right\rangle\left|6\right\rangle\\
	&S_{\mathrm{tot}}^{(6)}=2,S_{z,\mathrm{tot}}^{(6)}=0\\
	&\left|43\right\rangle=\left|2,0,\frac{1}{2},\frac{3}{2},0,1\right\rangle=
	\sqrt{\frac{1}{2}}\left|2\right\rangle\left|6\right\rangle+\sqrt{\frac{1}{2}}\left|1\right\rangle\left|7\right\rangle\\
	&\left|44\right\rangle=\left|2,0,\frac{1}{2},\frac{3}{2},1,1\right\rangle=
	\sqrt{\frac{1}{2}}\left|4\right\rangle\left|6\right\rangle+\sqrt{\frac{1}{2}}\left|3\right\rangle\left|7\right\rangle\\
	&\left|45\right\rangle=\left|2,0,\frac{3}{2},\frac{1}{2},1,0\right\rangle=
	\sqrt{\frac{1}{2}}\left|6\right\rangle\left|2\right\rangle+\sqrt{\frac{1}{2}}\left|7\right\rangle\left|1\right\rangle\\
	&\left|46\right\rangle=\left|2,0,\frac{3}{2},\frac{1}{2},1,1\right\rangle=
	\sqrt{\frac{1}{2}}\left|6\right\rangle\left|4\right\rangle+\sqrt{\frac{1}{2}}\left|7\right\rangle\left|3\right\rangle\\
	&\left|47\right\rangle=\left|2,0,\frac{3}{2},\frac{3}{2},1,1\right\rangle=
	\sqrt{\frac{1}{4}}\left|5\right\rangle\left|8\right\rangle+\sqrt{\frac{1}{4}}\left|6\right\rangle\left|7\right\rangle-
	\sqrt{\frac{1}{4}}\left|7\right\rangle\left|6\right\rangle-\sqrt{\frac{1}{4}}\left|8\right\rangle\left|5\right\rangle\\
	&S_{\mathrm{tot}}^{(6)}=2,S_{z,\mathrm{tot}}^{(6)}=1\\
	&\left|48\right\rangle=\left|2,1,\frac{1}{2},\frac{3}{2},0,1\right\rangle=
	\sqrt{\frac{1}{4}}\left|2\right\rangle\left|5\right\rangle+\sqrt{\frac{3}{4}}\left|1\right\rangle\left|6\right\rangle\\
	&\left|49\right\rangle=\left|2,1,\frac{1}{2},\frac{3}{2},1,1\right\rangle=
	\sqrt{\frac{1}{4}}\left|4\right\rangle\left|5\right\rangle+\sqrt{\frac{3}{4}}\left|3\right\rangle\left|6\right\rangle\\
	&\left|50\right\rangle=\left|2,1,\frac{3}{2},\frac{1}{2},1,0\right\rangle=
	\sqrt{\frac{1}{4}}\left|5\right\rangle\left|2\right\rangle+\sqrt{\frac{3}{4}}\left|6\right\rangle\left|1\right\rangle\\
	&\left|51\right\rangle=\left|2,1,\frac{3}{2},\frac{1}{2},1,1\right\rangle=
	\sqrt{\frac{1}{4}}\left|5\right\rangle\left|4\right\rangle+\sqrt{\frac{3}{4}}\left|6\right\rangle\left|3\right\rangle\\
	&\left|52\right\rangle=\left|2,1,\frac{3}{2},\frac{3}{2},1,1\right\rangle=
	\sqrt{\frac{1}{2}}\left|5\right\rangle\left|7\right\rangle-\sqrt{\frac{1}{2}}\left|7\right\rangle\left|5\right\rangle\\
	&S_{\mathrm{tot}}^{(6)}=2,S_{z,\mathrm{tot}}^{(6)}=2\\
	&\left|53\right\rangle=\left|2,2,\frac{1}{2},\frac{3}{2},0,1\right\rangle=\left|1\right\rangle\left|5\right\rangle\\
	&\left|54\right\rangle=\left|2,2,\frac{1}{2},\frac{3}{2},1,1\right\rangle=\left|3\right\rangle\left|5\right\rangle\\
	&\left|55\right\rangle=\left|2,2,\frac{3}{2},\frac{1}{2},1,0\right\rangle=\left|5\right\rangle\left|1\right\rangle\\
	&\left|56\right\rangle=\left|2,2,\frac{3}{2},\frac{1}{2},1,1\right\rangle=\left|5\right\rangle\left|3\right\rangle\\
	&\left|57\right\rangle=\left|2,2,\frac{3}{2},\frac{3}{2},1,1\right\rangle=
	\sqrt{\frac{1}{2}}\left|5\right\rangle\left|6\right\rangle-\sqrt{\frac{1}{2}}\left|6\right\rangle\left|5\right\rangle\\
	&S_{\mathrm{tot}}^{(6)}=3,S_{z,\mathrm{tot}}^{(6)}=-3,-2,-1,0,1,2,3\\
	&\left|58\right\rangle=\left|3,-3,\frac{3}{2},\frac{3}{2},1,1\right\rangle=\left|8\right\rangle\left|8\right\rangle\\
	&\left|59\right\rangle=\left|3,-2,\frac{3}{2},\frac{3}{2},1,1\right\rangle=\sqrt{\frac{1}{2}}\left|7\right\rangle\left|8\right\rangle+\sqrt{\frac{1}{2}}\left|8\right\rangle\left|7\right\rangle\\
	&\left|60\right\rangle=\left|3,-1,\frac{3}{2},\frac{3}{2},1,1\right\rangle=\sqrt{\frac{1}{5}}\left|6\right\rangle\left|8\right\rangle+\sqrt{\frac{3}{5}}\left|7\right\rangle\left|7\right\rangle+\sqrt{\frac{1}{5}}\left|8\right\rangle\left|6\right\rangle\\
	&\left|61\right\rangle=\left|3,0,\frac{3}{2},\frac{3}{2},1,1\right\rangle=\sqrt{\frac{1}{20}}\left|5\right\rangle\left|8\right\rangle+\sqrt{\frac{9}{20}}\left|6\right\rangle\left|7\right\rangle+\sqrt{\frac{9}{20}}\left|7\right\rangle\left|6\right\rangle+\sqrt{\frac{1}{20}}\left|8\right\rangle\left|5\right\rangle\\
	&\left|62\right\rangle=\left|3,1,\frac{3}{2},\frac{3}{2},1,1\right\rangle=\sqrt{\frac{1}{5}}\left|5\right\rangle\left|7\right\rangle+\sqrt{\frac{3}{5}}\left|6\right\rangle\left|6\right\rangle+\sqrt{\frac{1}{5}}\left|7\right\rangle\left|5\right\rangle\\
	&\left|63\right\rangle=\left|3,2,\frac{3}{2},\frac{3}{2},1,1\right\rangle=\sqrt{\frac{1}{2}}\left|5\right\rangle\left|6\right\rangle+\sqrt{\frac{1}{2}}\left|6\right\rangle\left|5\right\rangle\\
	&\left|64\right\rangle=\left|3,3,\frac{3}{2},\frac{3}{2},1,1\right\rangle=\left|5\right\rangle\left|5\right\rangle.
\end{align*}

\section{Uncompressed sequence}\label{app:sequence}

In this section, we provide the details of the uncompressed sequence obtained after applying the standard Krotov optimization, but prior to the application of the JK algorithm. This sequence successfully implements the desired Toffoli gate with high fidelity; however, it remains excessively long and inefficient, rendering it impractical for real-world quantum circuits. The uncompressed sequence serves as the starting point for the JK algorithm, which significantly compresses the pulse sequence while preserving gate fidelity. The details of the exchange unitaries constituting the sequence are shown in Table~\ref{tab:1-28} (for the first 28 time steps) and Table~\ref{tab:29-55} (for time steps 29 through 55). The optimization process is shown in Table~\ref{tab:convergence}.

\begin{table}[ht]
	\centering
	\renewcommand{\arraystretch}{1} 
	\setlength{\tabcolsep}{9pt} 
	\caption{55-time-step uncompressed sequence (from time step 1 to time step 28).}
	\begin{tabular}{|c|c|c|c|c|c|c|c|c|}
		\hline
		\text{Step} & $p_{1,2}$ & $p_{2,3}$ & $p_{3,4}$ & $p_{4,5}$ & $p_{5,6}$ & $p_{6,7}$ & $p_{7,8}$ & $p_{8,9}$ \\
		\hline
		1 & 0 & 0 & 0.171280 & 0 & 0.057610 & 0 & -0.113705 & 0 \\
		2 & 0 & -0.101342 & 0 & 0.177459 & 0 & 0 & 0 & 0.009875 \\
		3 & 0 & 0 & 0.328343 & 0 & 0.363308 & 0 & -0.075663 & 0 \\
		4 & 0 & 0.630905 & 0 & -0.609766 & 0 & 0 & 0 & 0.036130 \\
		5 & 0 & 0 & 0.391954 & 0 & 0.353324 & 0 & -0.012761 & 0 \\
		6 & 0 & -0.119032 & 0 & -0.362184 & 0 & -0.540480 & 0 & 0.015486 \\
		7 & 0 & 0 & 0.267867 & 0 & 0 & 0 & -0.015869 & 0 \\
		8 & 0 & -0.934082 & 0 & -0.046239 & 0 & 0 & 0 & -0.445718 \\
		9 & 0 & 0 & -0.821827 & 0 & 0.005940 & 0 & 0.775597 & 0 \\
		10 & 0 & -0.004373 & 0 & -0.547811 & 0 & 0 & 0 & -0.163575 \\
		11 & 0 & 0 & 0.668425 & 0 & 0.260181 & 0 & 0.004333 & 0 \\
		12 & 0 & 0.396335 & 0 & -0.817440 & 0 & 0 & 0 & -0.013776 \\
		13 & 0 & 0 & -0.441627 & 0 & -0.646837 & 0 & -0.048896 & 0 \\
		14 & 0 & 0.287595 & 0 & -0.434438 & 0 & 0 & 0 & -0.496229 \\
		15 & 0 & 0 & 0.952645 & 0 & 0.791044 & 0 & -0.880058 & 0 \\
		16 & 0 & 0.011442 & 0 & -0.544495 & 0 & 0 & 0 & -0.346432 \\
		17 & 0 & 0 & -0.506475 & 0 & -0.048677 & 0 & -0.081481 & 0 \\
		18 & 0 & -0.092802 & 0 & -0.476621 & 0 & 0 & 0 & -0.149473 \\
		19 & 0 & 0 & -0.542894 & 0 & -0.294417 & 0 & -0.666237 & 0 \\
		20 & 0 & 0.323522 & 0 & 0.129986 & 0 & 0.690990 & 0 & 0.102330 \\
		21 & 0 & 0 & -0.485848 & 0 & 0 & 0 & -0.567432 & 0 \\
		22 & 0 & -0.056718 & 0 & 0.487515 & 0 & -0.750763 & 0 & 0.597217 \\
		23 & 0 & 0 & -0.700015 & 0 & 0 & 0 & -0.897555 & 0 \\
		24 & 0 & 0.761959 & 0 & 0.797867 & 0 & -1.145150 & 0 & 0.503777 \\
		25 & 0 & 0 & 0.541105 & 0 & -1.383010 & 0 & 0.421025 & 0 \\
		26 & 0 & -0.094056 & 0 & 0 & 0 & -0.912389 & 0 & 0.391495 \\
		27 & 0 & 0 & -0.368176 & 0 & -1.434450 & 0 & -1.027200 & 0 \\
		28 & 0 & 0.858183 & 0 & 0 & 0 & 1.018070 & 0 & -0.627970 \\
		\hline
	\end{tabular}\label{tab:1-28}
\end{table}
\begin{table}[ht]
	\centering
	\renewcommand{\arraystretch}{1} 
	\setlength{\tabcolsep}{9pt} 
	\caption{55-time-step uncompressed sequence table (from time step 29 to time step 55).}
	\begin{tabular}{|c|c|c|c|c|c|c|c|c|}
		\hline
		\text{Step} & $p_{1,2}$ & $p_{2,3}$ & $p_{3,4}$ & $p_{4,5}$ & $p_{5,6}$ & $p_{6,7}$ & $p_{7,8}$ & $p_{8,9}$ \\
		\hline
		29 & 0 & 0 & 0.042789 & 0 & 1.001480 & 0 & 0.632431 & 0 \\
		30 & 0 & -0.024318 & 0 & 0 & 0 & 0.993828 & 0 & -0.671796 \\
		31 & 0 & 0 & -0.016895 & 0 & 0.937422 & 0 & -0.139507 & 0 \\
		32 & 0 & -0.564398 & 0 & 0 & 0 & 1.312100 & 0 & 0.140317 \\
		33 & 0 & 0 & -0.745372 & 0 & 0.503377 & 0 & -0.438745 & 0 \\
		34 & 0 & 0.013151 & 0 & 0 & 0 & 1.360290 & 0 & -0.521566 \\
		35 & 0 & 0 & -0.026124 & 0 & 0.693291 & 0 & 0.831985 & 0 \\
		36 & 0 & -0.382984 & 0 & 0 & 0 & -1.954380 & 0 & 0.395606 \\
		37 & 0 & 0 & -0.331504 & 0 & -0.470088 & 0 & -0.711540 & 0 \\
		38 & 0 & -0.343993 & 0 & 0 & 0 & -1.684190 & 0 & -0.250346 \\
		39 & 0 & 0 & -0.718306 & 0 & 0.709398 & 0 & 0.572356 & 0 \\
		40 & 0 & 0.595562 & 0 & 0 & 0 & 0.785922 & 0 & 0.598140 \\
		41 & 0 & 0 & 0.027358 & 0 & -0.132902 & 0 & -0.306381 & 0 \\
		42 & 0 & 0.320719 & 0 & -0.654041 & 0 & 0.513170 & 0 & 0.301276 \\
		43 & 0 & 0 & 0.391232 & 0 & -0.319153 & 0 & 0.862290 & 0 \\
		44 & 0 & 0.962250 & 0 & -0.133952 & 0 & 0.000011 & 0 & -0.104524 \\
		45 & 0 & 0 & -0.101004 & 0 & -0.277882 & 0 & 0.066079 & 0 \\
		46 & 0 & 0.443214 & 0 & -0.572575 & 0 & 0 & 0 & -0.037455 \\
		47 & 0 & 0 & 0.578033 & 0 & -0.378393 & 0 & -0.018994 & 0 \\
		48 & 0 & 0.857235 & 0 & 0.327071 & 0 & 0 & 0 & -0.230227 \\
		49 & 0 & 0 & -0.275722 & 0 & -1.084370 & 0 & -0.100310 & 0 \\
		50 & 0 & -0.171715 & 0 & -1.122270 & 0 & 0 & 0 & 0.051434 \\
		51 & 0 & 0 & -0.638390 & 0 & 0.033564 & 0 & -0.088825 & 0 \\
		52 & 0 & 0.011886 & 0 & -0.261533 & 0 & 0 & 0 & 0.537031 \\
		53 & 0 & 0 & 0.390378 & 0 & -0.241973 & 0 & 0.236710 & 0 \\
		54 & 0 & -0.424537 & 0 & 1.315600 & 0 & 0 & 0 & 0.428487 \\
		55 & 0 & 0 & -0.576510 & 0 & -0.132951 & 0 & -0.034890 & 0 \\
		\hline
	\end{tabular}\label{tab:29-55}
\end{table}

\begin{table}[ht]
	\centering
	\renewcommand{\arraystretch}{2} 
	\setlength{\tabcolsep}{10pt} 
	\begin{tabular}{|c|c|}
		\hline
		Number of iterations & Infidelity \\ \hline
		500     & $< 10^{-2}$ \\ \hline
		2,000   & $< 10^{-3}$ \\ \hline
		4,000   & $< 10^{-4}$ \\ \hline
		8,000   & $< 10^{-5}$ \\ \hline
		60,000  & $< 10^{-8}$ \\ \hline
	\end{tabular}
	\caption{Convergence of infidelity during the Krotov's method optimization process.}
	\label{tab:convergence}
\end{table}

\begingroup

\section{More gate and quantum circuit realized by JK algorithm}\label{app:others}

In addition to the Toffoli gate discussed in the main text, we further demonstrate the versatility of the JK algorithm by implementing other three qubit gates and circuits, including the Quantum Fourier Transform (QFT) and the Fredkin gate. The pulse sequences for these gates are optimized using the JK algorithm, and the accumulated numbers of exchange unitaries and time steps for each circuit are shown in the figures below.

The detailed implementation data, including the optimized pulse parameters for each gate and circuit, are available in our GitHub repository:

\url{https://github.com/WuJiahao555/JK-Algorithm-in-EO-Qubit}

The following figures illustrate the gate decomposition and the accumulated number of exchange unitaries and time steps for the QFT and Fredkin circuits:

\begin{figure}[h]
	\centering
	\includegraphics[width=0.8\textwidth]{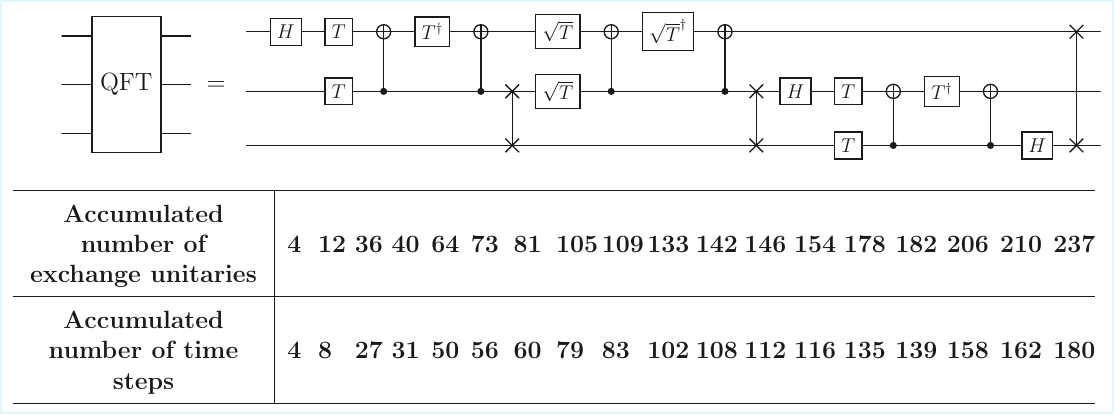}
	\caption{Gate decomposition and accumulated number of exchange unitaries and time steps for the three-qubit QFT circuit realized by the JK algorithm.}
	\label{fig:QFT3Circuit}
\end{figure}

\begin{figure}[h]
	\centering
	\includegraphics[width=0.8\textwidth]{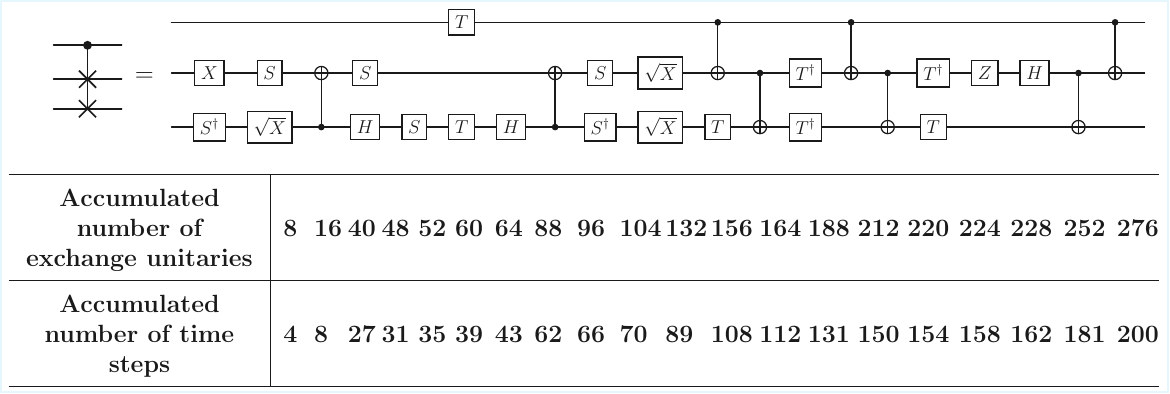}
	\caption{Gate decomposition and accumulated number of exchange unitaries and time steps for the Fredkin gate realized by the JK algorithm.}
	\label{fig:FredkinCircuit}
\end{figure}

When decomposed using conventional methods, the QFT circuit requires 180 time steps and 237 exchange unitaries, and the Fredkin gate requires 200 time steps and 276 exchange unitaries. In contrast, our JK algorithm was able to find pulse sequences that implement the same QFT circuit using only 80 time steps and 202 exchange unitaries, and the Fredkin gate using only 104 time steps and 172 exchange unitaries. These results further demonstrate the scalability and efficiency of the JK algorithm for synthesizing complex quantum gates and circuits with experimentally feasible pulse sequences.
\endgroup

\end{document}